\documentclass[paper,prd,reprint,nofootinbib,notitlepage,showpacs,showkeys]{revtex4-1}
\usepackage{graphicx}
\usepackage{amsmath}
\usepackage{amssymb}
\usepackage{dcolumn}% Align table columns on decimal point
\usepackage{bm}% bold math
\usepackage{color}
\usepackage{dsfont}

\newcommand \beq{\begin{eqnarray}}
\newcommand \eeq{\end{eqnarray}}
\newcommand \bml{\bar M^2_{\rm L}}
\newcommand \bmt{\bar M^2_{\rm T}}
\newcommand \bmz{\bar M^2_{\phi=0}}
\newcommand \bgl{\bar G_{\rm L}}
\newcommand \bgt{\bar G_{\rm T}}
\newcommand \bgz{\bar G_{\phi=0}}

\begin{document}
\allowdisplaybreaks

\title{Thermodynamics and phase transition of the $O(N)$ model\\ from the two-loop $\Phi$-derivable approximation}

\author{Gergely Mark{\'o}}
\email{smarkovics@hotmail.com}
\affiliation{Department of Atomic Physics, E{\"o}tv{\"o}s University, H-1117 Budapest, Hungary.}

\author{Urko Reinosa}
\email{reinosa@cpht.polytechnique.fr}
\affiliation{Centre de Physique Th{\'e}orique, Ecole Polytechnique, CNRS, 91128 Palaiseau Cedex, France.}

\author{Zsolt Sz{\'e}p}
\email{szepzs@achilles.elte.hu}
\affiliation{Statistical and Biological Physics Research Group of the Hungarian Academy of Sciences, H-1117 Budapest, Hungary.}

%\date{\today}

\begin{abstract}
We discuss the thermodynamics of the $O(N)$ model across the corresponding phase transition using the two-loop $\Phi$-derivable approximation of the effective potential and compare our results to those obtained in the literature within the Hartree-Fock approximation. In particular, we find that in the chiral limit the transition is of the second order, whereas it was found to be of the first order in the Hartree-Fock case. These features are manifest at the level of the thermodynamical observables. We also compute the thermal sigma and pion masses from the curvature of the effective potential. In the chiral limit, this guarantees that Goldstone's theorem is obeyed in the broken phase. A realistic parametrization of the model in the $N=4$ case, based on the vacuum values of the curvature masses, shows that a sigma mass of around 450~MeV can be obtained. The equations are renormalized after extending our previous results for the $N=1$ case by means of the general procedure described in Ref.~\cite{Berges:2005hc}. When restricted to the Hartree-Fock approximation, our approach reveals that certain problems raised in the literature concerning the renormalization are completely lifted. Finally, we introduce a new type of $\Phi$-derivable approximation in which the gap equation is not solved at the same level of accuracy as the accuracy at which the potential is computed. We discuss the consistency and applicability of these types of ``hybrid'' approximations and illustrate them in the two-loop case by showing that the corresponding effective potential is renormalizable and that the transition remains of the second order.
\end{abstract}


\maketitle 

%%%%%
\section{Introduction}

It is a well-known fact that the $\Phi$-derivable approximation scheme, also called in the literature two-particle irreducible (2PI) or Cornwall-Jackiw-Tomboulis (CJT) formalism, gives a first order phase transition when applied to the $O(N)$ model in its lowest, Hartree-Fock approximation level. Other resummation methods, such as the $1/N$ expansion give a second order phase transition already at leading order \cite{Petropoulos:1998gt}, in accordance with general expectations and with the result of the functional renormalization group approach \cite{Tetradis:1992xd,Ogure:1999xh}. It was argued \cite{Arnold:1992rz,Nemoto:1999qf} that close to the transition temperature the contribution of higher loops may become important and that already the inclusion of the setting-sun diagram in the $\Phi$-derivable functional will render the phase transition of the second order type. 

As a continuation of our previous investigation done for the one-component scalar model, where we found that the change of order indeed happens within a full two-loop treatment of the effective action, we turn now to the physically more interesting $O(N)$ model. For $N=4$ this model can be regarded as a low energy effective model of two flavor QCD because the global $SU(2)_{\rm L}\times SU(2)_{\rm R}$ symmetry of the latter is isomorphic with $O(4).$ Since the $O(4)$ model contains both the longitudinal and transverse excitations of the chiral order parameter, it is widely used in the phenomenology of low energy mesons for the qualitative description of medium induced effects, especially around the phase transition. 

We would like to understand where exactly the contribution coming from the setting-sun diagram is essential to obtain the right order of the phase transition. Therefore, in addition to the two-loop approximation, we consider an approximation where the effective action is computed at two-loop order, but is evaluated for propagators computed from the Hartree-Fock approximation. Although this hybrid type of approximation might present certain inconsistencies, as it does not obey the conditions identified by Baym in Ref.~\cite{Baym:1962sx}, it is convenient in practice because its numerical treatment is much easier. Moreover, we will see that, for not too low temperatures where the hybrid approximation does not seem to show inconsistencies, its results are pretty close to those obtained from the two-loop approximation which is numerically more time and memory demanding. In particular, both in the two-loop and in the hybrid approximations, the transition is found to be of the second order.

As far as meson phenomenology is concerned, we will be particularly interested in the value of the sigma mass which can be obtained within a realistic parametrization of the model. Indeed, one of the difficulties when applying the $O(4)$ model to meson phenomenology is to obtain high enough values of the sigma mass, while maintaining the interpretation as an effective model where the cutoff $\Lambda,$ a mere separation scale between the modes of interest ($p\ll\Lambda$) and those which are integrated out ($p>\Lambda$), does not play the role of a parameter. This is usually rendered difficult by the fact that the model possesses a Landau pole in the ultraviolet and, if the latter is too close to the physical scales, the renormalization procedure is not enough to ensure the insensitivity of the results to cutoff values below the Landau pole. We will see that, in the two-loop and hybrid approximations, one can obtain reasonable values of the sigma mass with a Landau scale almost one order of magnitude higher. This, combined with the fact that the Landau pole does not show up in the renormalized quantities defined within the two-loop or hybrid approximations, allows us to meet the above mentioned requirements.

In principle the insensitivity to the cutoff scale $\Lambda$ is ensured automatically by the renormalization group since, by following a line of constant physics, a change in $\Lambda$ is carried over to the (bare) parameters of the Lagrangian, in such a way that the low energy physics is unaffected. Even if this  picture persists order by order in perturbation theory, this is not necessarily so for approximation schemes that go beyond it and certain amendments need to be made to the renormalization procedure, depending on the method used. Over the last few years a general method for renormalizing $\Phi$-derivable approximations has been developed and we illustrate it here both in the two-loop and in the hybrid approximations to the 2PI effective action. By revisiting the lower Hartree-Fock approximation, we can also compare our renormalization procedure to other approaches followed in the literature. In particular, we show that certain inconsistencies discussed in Ref.~\cite{Lenaghan:1999si} are completely lifted by our approach.

In Secs.~\ref{sec:approximations} and \ref{sec:hybrid}, we define and renormalize the two approximations to be discussed in this work and compare our renormalization procedure to other approaches. Section \ref{sec:numerics} is devoted to some of the numerical tricks that we use to achieve high accuracy results in the two-loop approximation. Section \ref{sec:results} deals with the parametrization of the model, with a special attention to the attainable values of sigma mass and gathers our results on the phase transition and thermodynamical observables. We also discuss there the dependence of the physical quantities on the renormalization scale and on the cutoff. We present our conclusions in Sec.~\ref{sec:conclusions}.\\

%%%%%
\section{Two-loop approximation}\label{sec:approximations}

%%%
\subsection{Relevant equations}\label{subsec:equations}

The 2PI effective action for the $O(N)$ model is a functional of a one-point function $\phi_a(x)$ and a symmetric two-point function $G_{ab}(x,y)=G_{ba}(y,x)$. In the imaginary-time formulation of field theory at a finite temperature $T=1/\beta$ and at two-loop order, it reads
\begin{widetext}
\beq\label{eq:2PI}
\Gamma[\phi,G] & = & \frac{m_2^2}{2}\,\int_x\phi^2(x)+\frac{\lambda_4}{24N}\,\int_x(\phi^2(x))^2+\frac{1}{2}\int_x{\rm tr}\,\big[\ln G^{-1}+(-\square_E+m^2_0)\,G-1\big](x,x)\nonumber\\
& & +\,\frac{\lambda_2^{(A)}}{12N}\,\int_x\phi^2(x)\,{\rm tr}\,G(x,x)+\frac{\lambda_2^{(B)}}{6N}\,\int_x\phi(x)G(x,x)\phi(x)+\frac{\lambda_0^{(A)}}{24N}\,\int_x [{\rm tr}\,G(x,x)]^2+\frac{\lambda_0^{(B)}}{12N}\,\int_x {\rm tr}\,G^2(x,x)\nonumber\\
& & -\,\frac{\lambda^2_\star}{36N^2}\int_x\int_y\phi(x)G(x,y)\phi(y)\,{\rm tr}\,[G(x,y)G(y,x)]-\frac{\lambda^2_\star}{18N^2}\int_x\int_y\phi(x)G(x,y)G(y,x)G(x,y)\phi(y)\,,
\eeq
\end{widetext}
with $\int_x\equiv\int_0^\beta d\tau\int d^3x$, $\phi^2\equiv\phi_a\phi_a$, $\phi G\phi\equiv \phi_aG_{ab}\phi_b$, ${\rm tr}\,G\equiv G_{aa}$ and where a summation over repeated indices is implied. As discussed in Refs.~\cite{Berges:2005hc,Marko:2012wc} and below, the need for two bare masses $m_0$ and $m_2$ and three families of bare couplings, labeled with the indices $0$, $2$ and $4$ respectively, reflects the fact that, given a truncation of the 2PI effective action, there are two possible ways to define the two-point function and three different ways to define the four-point function. As we discuss in Appendix~\ref{app:four-point}, the need for a doubling (represented by the superscripts $A$ and $B$) of the bare couplings carrying an index $0$ or $2$ has to do with the fact that two of these four-point functions do not obey the crossing symmetry.\footnote{Equivalently, the corresponding terms in the 2PI effective action (\ref{eq:2PI}) are independently invariant under $O(N)$ transformations, see Ref.~\cite{Berges:2005hc}.}  Finally, following Ref.~\cite{Marko:2012wc}, we have replaced the bare couplings in the highest loop diagrams of Eq.~(\ref{eq:2PI}) by a coupling $\lambda_\star$ which will be identified later to the renormalized coupling at some renormalization scale $T_\star$. This is because no renormalization comes from these vertices at this level of truncation.\\

In what follows, we study the phase transition of the model by computing the effective potential $\gamma(\phi)$. The latter is obtained after evaluating the functional (\ref{eq:2PI}) at the stationary value of $G$ which we denote $\bar G_\phi$,\footnote{In order to alleviate the notations, the dependence of $\bar G$ and alike on various quantities such as $\phi$, $T$, \dots will be written explicitly only when needed.} with $\phi$ a homogeneous field configuration:
\beq
\gamma(\phi)=\frac{1}{\beta V}\Gamma[\phi,\bar G_\phi]\,.
\label{Eq:gamma_1PI}
\eeq
Some more explicit expressions of the effective potential will be given later. In the presence of a homogeneous field, the propagator $\bar G(x,y)$ depends on the difference $x-y$ or, in Fourier space, on $Q=(i\omega_n,\vec{q})$ where $\omega_n=2\pi n/\beta$ is a bosonic Matsubara frequency. From parity and time-reversal symmetry, we have $\bar G_{ab}(Q)=\bar G_{ab}(-Q)$ and thus $\bar G_{ab}(Q)=\bar G_{ba}(Q)$ since $\bar G_{ab}(Q)=\bar G_{ba}(-Q)$. Moreover, the $O(N)$ invariance of $\Gamma[\phi,G]$ upon simultaneous rotation of $\phi$ and $G$ implies that $\bar G_\phi$ is covariant:
\beq\label{eq:cov}
\bar G^{R\phi}_{ab}=R_{ac}R_{bd}\bar G^\phi_{cd}\,, \quad \forall\, R\in O(N)\,.
\eeq
We recall in Appendix~\ref{app:tensor} that, together with the property $\bar G_{ab}(Q)=\bar G_{ba}(Q)$, this implies the following spectral decomposition
\beq\label{eq:tensor}
\bar G_{ab}=\bar G_{\rm L} P^{\rm L}_{ab}+\bar G_{\rm T}P^{\rm T}_{ab}\,,
\label{Eq:prop_projected}
\eeq
with
\beq
P^{\rm L}_{ab}\equiv\frac{\phi_a\phi_b}{\phi^2} \quad {\rm and} \quad P^{\rm T}_{ab}\equiv\delta_{ab}-\frac{\phi_a\phi_b}{\phi^2}
\label{Eq:projectors}
\eeq
the longitudinal and transverse projectors with respect to $\phi$ and where the functions $\bar G_{\rm L}$ and $\bar G_{\rm T}$ depend on $\phi$ only through $\phi^2$. 

It is convenient to introduce momentum dependent longitudinal and transverse masses defined through the relation $\bar G_{\rm L,T}(Q)=1/(Q^2+\bar M^2_{\rm L,T}(Q)),$ such that they include the corresponding self-energy and the tree-level mass. After some straightforward calculation starting from the stationarity condition $0=\delta\Gamma/\delta G|_{\bar G}$, one shows using the two projectors in Eq.~\eqref{Eq:projectors} that they obey the following coupled gap equations:\footnote{By using the substitutions $\lambda_{0,2}^{\alpha A+\beta B}/(\alpha+\beta)\to 12/\varepsilon$ and $\lambda_\star\to 12/\varepsilon$, we recover the equations derived in Ref.~\cite{Seel:2011ju}. There however, the equations were neither renormalized nor solved.}
\beq\label{eq:bml}
\bml(K) & = & m^2_0+\frac{\lambda^{(A+2B)}_0}{6N}{\cal T}[\bgl]+\frac{\lambda^{((N-1)A)}_0}{6N}{\cal T}[\bgt]\nonumber\\
&&-\frac{\lambda_\star^2}{18N^2}\phi^2\big[9{\cal B}[\bgl](K)
+(N-1){\cal B}[\bgt](K)\big]\nonumber\\
& & +\frac{\lambda^{(A+2B)}_2}{6N}\phi^2,
%& & +\,\frac{\phi^2}{6N}\left[\lambda^{(A+2B)}_2-\frac{3\lambda^2_\star}{N}{\cal B}[\bgl](K)\right.\nonumber\\
%& & \hspace{2.4cm}\left.-\,\frac{N-1}{3N}\lambda^2_\star{\cal B}[\bgt](K)\right]\!,
\eeq
and
\beq\label{eq:bmt}
\bmt(K) & = & m^2_0+\frac{\lambda^{(A)}_0}{6N}{\cal T}[\bgl]+\frac{\lambda^{((N-1)A+2B)}_0}{6N}{\cal T}[\bgt]\nonumber\\
& & +\,\frac{\phi^2}{6N}\left[\lambda^{(A)}_2-\frac{2\lambda^2_\star}{3N}{\cal B}[\bgl;\bgt](K)\right]\!.
\eeq
In order to save space, we find it appropriate to write 
\beq
\lambda_{0,2}^{(\alpha A+\beta B)}\equiv \alpha\lambda_{0,2}^{(A)}+\beta\lambda_{0,2}^{(B)}\,,
\eeq
and denoting the sum integral by
\beq
\int_Q^T f(Q)\equiv T\sum_{n=-\infty}^{\infty}\int\frac{d^3q}{(2\pi)^3} f(i\omega_n,q),
\eeq
where $q=|\vec{q}|$, we use the short-hand notations
\beq
{\cal T}[G] & \equiv & \int_Q^T G(Q)\,,\\
{\cal B}[G_1;G_2](K) & \equiv & \int_Q^T G_1(Q)G_2(Q+K)\,,\\
{\cal S}[G_1;G_2;G_3] & \equiv & \int_K^T\int_Q^T G_1(K)G_2(Q)G_3(Q+K)\,.\nonumber\\
\eeq
For the last two of them, when all the arguments are equal to a given propagator $G$, we write more simply ${\cal B}[G](K)$ and ${\cal S}[G]$. Taking the difference of Eqs.~(\ref{eq:bml}) and (\ref{eq:bmt}), it is straightforward to check that, when $\phi=0$, the system of equations is compatible with a solution such that $\bml=\bmt\equiv\bar M^2_{\phi=0}$ with
\beq\label{eq:bmz}
\bmz= m^2_0+\frac{\lambda^{(NA+2B)}_0}{6N}{\cal T}[\bgz]
\eeq
and $\bar G_{\phi=0}(Q)\equiv 1/(Q^2+\bar M^2_{\phi=0})$. 

The nature of the transition will be discussed by monitoring the nontrivial extrema $\bar\phi$ of the effective potential. They obey the equation
\beq\label{eq:field}
0 & = & m^2_2+\frac{\lambda_4}{6N}\bar\phi^2+\frac{\lambda_2^{(A+2B)}}{6N}{\cal T}[\bgl]+\frac{\lambda_2^{((N-1)A)}}{6N}{\cal T}[\bgt]\nonumber\\
& & -\frac{\lambda^2_\star}{18N^2}\Big(3{\cal S}[\bgl]+(N-1){\cal S}[\bgl;\bgt;\bgt]\Big),
\eeq
which, due to the stationarity condition $0=\delta\Gamma/\delta G|_{\bar G}$, originates only from the explicit field dependence of the functional (\ref{eq:2PI}). We note that the case $N=1$ is obtained after disregarding Eq.~(\ref{eq:bmt}) and making the replacements $\lambda_{0,2}^{(NA+2B)}=\lambda_{0,2}^{(A+2B)}\to 3\lambda_{0,2}$ in Eqs.~(\ref{eq:bml}) and (\ref{eq:field}). We shall use this recipe later in order to cross-check the expressions obtained for the bare parameters.\\ 

We shall also need the curvature of the potential, which at $\phi=0$ is found to be
\beq\label{eq:curvature}
\hat M^2_{\phi=0}=m^2_2+\frac{\lambda_2^{(NA+2B)}}{6N}{\cal T}[\bgz]-\frac{N+2}{18N^2}\lambda^2_\star{\cal S}[\bgz]\,.\nonumber\\
\eeq
More generally, one can define the curvature tensor at an arbitrary value of the field. Using as in Ref.~\cite{Aarts:2002dj} that the effective potential depends on the field $\phi$ only through the $O(N)$-invariant $\phi^2$, one writes $\gamma(\phi)=U(\phi^2)$ and obtains
\beq
\nonumber
\hat M_{ab}(\phi)&=&\frac{\delta^2\gamma(\phi)}{\delta\phi_a \delta\phi_b}=4U''(\phi^2)\phi_a\phi_b+2U'(\phi^2)\delta_{ab}\\
&=&\big[2U'(\phi^2)+4\phi^2 U''(\phi^2)\big]P_{ab}^{\rm L}+2U'(\phi^2)P_{ab}^{\rm T}\,.\nonumber\\ 
\eeq
In this paper we shall call curvature masses the two eigenmodes appearing in the above equation, evaluated at the solution $\bar \phi$ of the field equation:
\beq
\hat M^2_{\rm L}=2U'(\bar\phi^2)+4\phi^2 U''(\bar\phi^2) \,\, {\rm and} \,\, \hat M^2_{\rm T}=2U'(\bar\phi^2).
\label{Eq:curvature_masses}
\eeq
The field equation reads
\beq
\frac{\delta\gamma(\phi)}{\delta \phi_a}\bigg|_{\phi=\bar\phi}=2U'(\bar\phi^2)\,\bar\phi_a=0\,,
\eeq
from which it follows first, that $\hat M^2_{\rm L}=\hat M^2_{\rm T}$ in the symmetric phase (since $\bar\phi=0$) and second, that $\hat M^2_{\rm T}=0$ in the broken phase (since $\bar\phi\neq 0$ and thus $U'(\bar\phi^2)=0$) in agreement with Goldstone's theorem. In contrast, there is no reason for the gap mass $\bar M^2_{\rm T}\equiv\bmt(K=0)$ to vanish in the broken phase and we shall investigate quantitatively how much the Goldstone's theorem is violated in this case.

In the case of explicitly broken symmetry, when a term $-h\phi\equiv -h_a\phi_a$ is added to the effective potential, what changes is the field equation,\footnote{The stationarity condition that defines $\bar G_\phi$ is not changed. It follows that Eq.~(\ref{eq:cov}) and in turn Eq.~(\ref{eq:tensor}) still hold.} which becomes $\delta \gamma(\phi)/\delta \phi_a|_{\bar\phi}=2U'(\bar\phi^2)\bar\phi_a-h_a=0,$ so that we have $\hat M^2_{\rm T}=||h||/||\bar\phi||$ and $\hat M^2_{\rm L}$ still given by Eq.~\eqref{Eq:curvature_masses}. Without loss of generality, we can choose $h=(||h||,0,\dots,0)$ along the first coordinate axis. Note also that if we view $2U'(\phi^2)=f(||\phi||)$ as a function of $||\phi||$, we can compute the longitudinal curvature mass as $\hat M^2_{\rm L}=f(||\phi||)+||\phi||f'(||\phi||)$ from a numerical derivative of the function $f(||\phi||)$ which appears on the right-hand side of Eq.~(\ref{eq:field}).\\

The gap and field equations (\ref{eq:bml}), (\ref{eq:bmt}) and (\ref{eq:field}) will be solved using the techniques developed in Ref.~\cite{Marko:2012wc} that we quickly summarize in Sec.~\ref{sec:numerics}. Before we proceed to the numerical resolution of the model, we must however determine the values of the bare parameters in such a way that the sensitivity to the ultraviolet regulator is removed, or at least considerably reduced. The results that we shall present are valid for any regularization that can be defined nonperturbatively. For definiteness however and in line with the numerical method that we use to solve the two-loop approximation, in the next section, we assume that 3D momenta of modulus larger than a given cutoff $\Lambda$ are dropped. More details concerning the regularization procedure can be found in Ref.~\cite{Marko:2012wc}.

%%%
\subsection{Renormalization}\label{subsec:renormalization}
As explained in Ref.~\cite{Berges:2005hc} and illustrated in Ref.~\cite{Marko:2012wc}, the fact that the gap masses at zero momentum are different from the curvature masses requires the presence of two distinct bare masses $m_0$ and $m_2$. Those are fixed by means of the usual renormalization condition 
\beq
\bar M^2_{\phi=0,T_\star}(K=0)=m^2_\star
\eeq
at some renormalization scale, here a temperature $T_\star$, supplemented by a consistency condition
\beq
\hat M^2_{\phi=0,T_\star}=\bar M^2_{\phi=0,T_\star}(K=0)\,,
\eeq 
which restores the equality of the two masses at the renormalization point. Applying these conditions to Eqs.~(\ref{eq:bmz}) and (\ref{eq:curvature}), we obtain
\beq\label{Eq:m0_explicit}
m^2_0=m^2_\star-\frac{\lambda^{(NA+2B)}_0}{6N}{\cal T}_\star[G_\star]
\eeq
and
\beq\label{Eq:m2_explicit}
m^2_2=m^2_\star-\frac{\lambda^{(NA+2B)}_2}{6N}{\cal T}_\star[G_\star]+\frac{N+2}{18N^2}\,\lambda^2_\star{\cal S}_\star[G_\star]\,,\ \ \ 
\eeq
with $G_\star(Q_\star)\equiv 1/(Q_\star^2+m^2_\star)$. The $\star$ on any quantity means that it is computed at the temperature $T_\star$. For instance, $Q_\star$ means that the corresponding Matsubara frequencies involve the temperature $T_\star$. Similar considerations apply to the four-point function which admits three distinct definitions $\bar V_{ab,cd}$, $V_{ab,cd}(K)$, $\hat V_{abcd}$; see Appendix~\ref{app:four-point}. The first two do not obey the crossing symmetry and thus involve two independent components at $\phi=0$: $\bar V^{\phi=0}_{ab,cd}=\bar V^{(A)}_{\phi=0}\delta_{ab}\delta_{cd}+\bar V^{(B)}_{\phi=0}(\delta_{ac}\delta_{bd}+\delta_{ad}\delta_{bc})$ and similarly for $V^{\phi=0}_{ab,cd}(K)$. In contrast $\hat V^{\phi=0}_{abcd}=\hat V_{\phi=0}(\delta_{ab}\delta_{cd}+\delta_{ac}\delta_{bd}+\delta_{ad}\delta_{bc})$ is crossing symmetric. The renormalization condition 
\beq
\bar V^{(A)}_{\phi=0,T_\star}=\frac{\lambda_\star}{3N}
\eeq
and the consistency conditions 
\beq
\hat V_{\phi=0,T_\star} & = & \bar V^{(A)}_{\phi=0,T_\star}=\bar V^{(B)}_{\phi=0,T_\star}\nonumber\\
& = & V^{(A)}_{\phi=0,T_\star}(K=0)=V^{(B)}_{\phi=0,T_\star}(K=0)
\eeq
fix all the bare couplings that we have introduced and restore the equality of the various four-point functions at the renormalization point, in particular, the crossing symmetry becomes manifest. We obtain the following expressions for the bare parameters:
\beq\label{eq:l02B}
\frac{3N}{\lambda^{(B)}_0}=\frac{3N}{\lambda_\star}-{\cal B}_\star[G_\star](0)
\eeq
and
\beq\label{eq:l0NA2B}
\frac{3N}{\lambda^{(NA+2B)}_0}=\frac{3N}{(N+2)\lambda_\star}-\frac{1}{2}{\cal B}_\star[G_\star](0)\,,
\eeq
for those coupling parameters labeled with $0$, $\lambda_2^{(B)}\equiv\lambda^{(B)}_{2{\rm l}}+\delta\lambda^{(B)}_{2{\rm nl}}$ with
\beq
\label{eq:dl2Bnl}
\delta\lambda^{(B)}_{2{\rm nl}}=\frac{N+6}{6N}\,\lambda^2_\star\,{\cal B}_\star[G_\star](0)
\eeq
and
\beq
\frac{\lambda^{(B)}_{2{\rm l}}}{\lambda^{(B)}_0}=1-\frac{N+6}{18N^2}\,\lambda^2_\star\int_{Q_\star}^{T_\star}G^2_\star(Q_\star)\Delta{\cal B}_\star[G_\star](Q_\star)\,,\ \ 
\eeq
as well as $\lambda_2^{(NA+2B)}\equiv\lambda^{(NA+2B)}_{2{\rm l}}+\delta\lambda^{(NA+2B)}_{2{\rm nl}}$ with
\beq
\label{eq:dl2NAp2Bnl}
\delta\lambda^{(NA+2B)}_{2{\rm nl}}=\frac{N+2}{N}\lambda^2_\star\,{\cal B}_\star[G_\star](0)
\eeq
and
\beq\label{eq:l2lNA2B}
\frac{\lambda^{(NA+2B)}_{2{\rm l}}}{\lambda^{(NA+2B)}_0}=1-\frac{N+2}{6N^2}\lambda^2_\star\int_{Q_\star}^{T_\star}G^2_\star(Q_\star)\Delta{\cal B}_\star[G_\star](Q_\star)\,,\nonumber\\
\eeq
for those coupling parameters labeled with $2$ and finally
\beq
\lambda_4 & = & -2\lambda_\star+\frac{1}{N}\frac{\big(\lambda^{(NA+2B)}_{2{\rm l}}\big)^2}{\lambda^{(NA+2B)}_0}+2\left(1-\frac{1}{N}\right)\frac{\big(\lambda^{(B)}_{2{\rm l}}\big)^2}{\lambda^{(B)}_0}\nonumber\\
& & +\,\lambda^4_\star\left[\frac{(N+2)^2}{6N^4}+\frac{(N+6)^2}{54N^3}\left(1-\frac{1}{N}\right)\right]\nonumber\\
& & \times\int_{Q_\star}^{T_\star}G^2_\star(Q_\star)\left[\Delta{\cal B}_\star[G_\star](Q_\star)\right]^2\,.
\eeq
In the above expressions, $\Delta {\cal B}_\star[G_\star](Q_\star)$ stands for the difference of bubble sum integrals ${\cal B}_\star[G_\star](Q_\star)-{\cal B}_\star[G_\star](0)$.  The reason for the splitting of the bare parameters $\lambda_2^{(A,B)}$ into ``local'' and ``nonlocal'' parts, $\lambda_{2{\rm l}}^{(A,B)}$ and $\delta\lambda_{2{\rm nl}}^{(A,B)}$ respectively, is explained in Refs.~\cite{Reinosa:2003qa,Fejos:2011zq,Reinosa:2011cs}; see also Appendix~\ref{app:hybrid}. Applying the replacement rule discussed right after Eq.~(\ref{eq:field}), one recovers the $N=1$ bare parameters of Ref.~\cite{Marko:2012wc}. It is also simple to obtain the expressions for the bare parameters in the Hartree-Fock approximation. One has simply to set $\delta\lambda_{2{\rm nl}}^{(A,B)}=0$ and to remove all those terms that involve $\Delta {\cal B}_\star[G_\star]$. Then $\lambda^{(A)}_0$ and $\lambda^{(B)}_0$ remain unchanged, while $m_2=m_0$, $\lambda^{(A)}_2=\lambda^{(A)}_0$, $\lambda^{(B)}_2=\lambda^{(B)}_0$ and
\beq\label{eq:l4H}
\lambda_4^{\rm H}=-2\lambda_\star+\lambda^{(A+2B)}_0\,,
\eeq
which gives $\lambda_4^{\rm H}=-2\lambda_\star+3\lambda_0$ when $N=1$, in agreement with the result of Ref.~\cite{Reinosa:2011ut}. We have introduced a superscript `H' for the value taken by $\lambda_4$ in the Hartree-Fock approximation for later convenience.\\

Following similar steps as in Ref.~\cite{Marko:2012wc}, it is possible to prove implicitly that the bare parameters given above renormalize the gap and field equations, as well as the effective potential (up to a temperature and field independent divergence for this latter quantity). By ``implicit proof'', we mean that certain steps require some assumptions on the properties of a function, the spectral function, which is defined implicitly. We are not able to prove these properties but we can argue that they are plausible for they are true perturbatively and the resummation should only bring innocuous logarithmic corrections to them. We shall not reproduce this proof here and refer to Ref.~\cite{Marko:2012wc} for further details. In the next section however, we illustrate some of the aspects of the proof which are specific to the $O(N)$ model by using a simpler approximation where renormalization can be performed in an explicit way. This will be also the opportunity to revisit the renormalization of the Hartree-Fock approximation from our point of view and to compare to other results in the literature, in particular those of Ref.~\cite{Lenaghan:1999si}.\\

%%%
\subsection{Landau pole}
Let us end this section by discussing the presence of a Landau pole in the $O(N)$ model and how this affects the discussion of renormalization at the level of approximation considered in this work.\\

First of all, at least one pole is present in the expressions for the bare parameters.  Indeed, the equations (\ref{eq:l02B}) and (\ref{eq:l0NA2B}) determining the bare couplings $\lambda_0^{(A)}$ and $\lambda_0^{(B)}$ can be rewritten as
\beq
\frac{1}{\lambda_0^{(B)}}=\frac{1}{\lambda_\star}\left[1-\frac{2\lambda_\star}{6N}{\cal B}_\star^\Lambda[G_\star](0)\right]
\eeq
and
\beq
\frac{1}{\lambda_0^{(A)}}=\frac{1}{\lambda_0^{(B)}}\left[1-\frac{(N+2)\lambda_\star}{6N}{\cal B}_\star^\Lambda[G_\star](0)\right],
\eeq
where we have made the cutoff dependence of the bubble sum integral explicit. Since the latter grows logarithmically with $\Lambda$, it follows that both $\lambda_0^{(A)}$ and $\lambda_0^{(B)}$ diverge before turning negative at some value of $\Lambda$, which signals an instability. The bare coupling $\lambda_0^{(A)}$ being the first to diverge since $N>0$, it is natural to define the Landau scale $\Lambda_{\rm p}$ from the equation:
\beq\label{Eq:Lp}
0=1-\frac{N+2}{6N}\lambda_\star {\cal B}_\star^{\Lambda_{\rm p}}[G_\star](0)\,.
\eeq
Above this scale, at least one of the bare couplings becomes negative and one might wonder whether the theory is stable. In contrast, below this scale, it is easily checked, using the fact that ${\cal B}_\star[G_\star](0)>0$ and $\Delta {\cal B}_\star[G_\star](Q_\star)<0$ (this is proven for instance in Appendix~B.3 of Ref.~\cite{Marko:2012wc}), that all the bare couplings remain positive. To remain in the stability region, we shall thus consider values of $\Lambda$ below $\Lambda_{\rm p}$.\\

In the case of the two-loop approximation considered here (and also in the hybrid approximation that we introduce in the next section or in the Hartree-Fock approximation), the presence of a pole in the cutoff dependence of the bare couplings does not imply the appearance of a pole in the integrals that enter the physical observables. Choosing parameters such that the Landau scale is not too close to the physical scales,\footnote{If the Landau scale is too close to the other scales, we have seen in Ref.~\cite{Reinosa:2011cs} that the gap equation might lose its solution if the cutoff is taken too large, implying that the physical observables are not defined for too large values of the cutoff. But this is not due to the appearance of a pole in the integrals contributing to these observables.} the physical quantities are defined for any value of the cutoff $\Lambda$. This is because in the two-loop approximation (and also in the Hartree-Fock approximation or in the hybrid approximation considered in the next section) the self-energy does not grow quadratically at large frequency/momentum and also because these approximations do not involve vertex-type resummations capable of generating a Landau pole in the physical quantities. It follows that one can discuss renormalization as usual, in terms of divergent and convergent quantities as $\Lambda\rightarrow\infty$ and thus, even though we restrict to values of $\Lambda$ below $\Lambda_{\rm p}$, the renormalization procedure ensures that the results are already pretty insensitive to the cutoff in this range if the Landau scale is large enough. We have already studied these features in Refs.~\cite{Reinosa:2011ut,Marko:2012wc} and we shall also do it here briefly in Sec.~\ref{sec:results}. 

At higher orders of approximation, one expects a pole to appear in the physical observables too, at a finite value of the cutoff. This prevents discussing the renormalization in terms of divergent and convergent quantities as $\Lambda\to\infty$. Still, if the Landau scale is large enough, these concepts survive in a somewhat generalized acceptation. In particular, quantities renormalized according to our scheme will still show a plateau behavior below the Landau scale, from which one can extract results that are pretty insensitive to the cutoff. The discussion becomes more delicate as the Landau scale gets closer to the physical scales.

%%%%%
\section{Hybrid approximation}\label{sec:hybrid}
We shall also consider another type of approximation where the gap equation is solved at a lower level of accuracy than that used to compute the effective potential. We name these approximations ``hybrid'' for they break to some extent the consistency of the $\Phi$-derivable formalism. In particular, because the potential is not evaluated at its stationary point, the field equation admits additional contributions of the form $\delta\Gamma/\delta G|_{\bar G}\,\delta\bar G/\delta\phi$. These types of approximations have been considered in earlier works as well; see \cite{Bordag:2000tb,Bordag:2001jf,Arrizabalaga:2002hn}. We note that these types of approximations do not obey Baym's conditions and might thus lead to certain inconsistencies in some region of the parameters.\\

%%%
\subsection{Definition and relevant equations}
To make things explicit, we consider the two-loop 2PI effective potential (below $c_{\rm L}=1$ and $c_{\rm T}=N-1$):
%\begin{widetext}
%\beq
%\label{eq:2loop_effpot}
%\gamma[\phi,G_{\rm L},G_{\rm T}] & = & N\gamma_0(m_\star)+\frac{1}{2}\sum_{i={\rm T},{\rm L}}c_i\int_Q^T \big[\ln G^{-1}_i(Q)-\ln G_\star^{-1}(Q)+(Q^2+m^2_0)\,G_i(Q)-1\big]\nonumber\\
%&& +  \frac{1}{2}m_2^2\phi^2+\frac{\lambda_4}{24N}\phi^4+\frac{\lambda_2^{(A+2B)}}{12N}\phi^2{\cal T}[G_{\rm L}]+\frac{\lambda_2^{((N-1)A)}}{12N}\phi^2{\cal T}[G_{\rm T}]\nonumber\\
%&& +  \frac{\lambda_0^{(A+2B)}}{24N}\,{\cal T}^2[G_{\rm L}]+\frac{\lambda_0^{((N-1)A)}}{12N}{\cal T}[G_{\rm L}]{\cal T}[G_{\rm T}]+\frac{\lambda_0^{((N-1)^2 A+2(N-1)B)}}{24N}\,{\cal T}^2[G_{\rm T}]\nonumber\\
%&& -  \frac{\lambda^2_\star}{12N^2}\phi^2{\cal S}[G_{\rm L}]-\frac{(N-1)\lambda^2_\star}{36N^2}\phi^2{\cal S}[G_{\rm L};G_{\rm T};G_{\rm T}]\,,
%\eeq
%\end{widetext}
\begin{widetext}
\beq
\label{eq:2loop_effpot}
\gamma[\phi,G_{\rm L},G_{\rm T}] & = & N\gamma_0(m_\star) +  \frac{1}{2}m_2^2\phi^2+\frac{\lambda_4}{24N}\phi^4
+\sum_{i={\rm T},{\rm L}}\frac{c_i}{2}\int_Q^T \big[\ln G^{-1}_i(Q)-\ln G_\star^{-1}(Q)+(Q^2+m^2_0)\,G_i(Q)-1\big]\nonumber\\
&& +  \frac{\lambda_0^{(A+2B)}}{24N}\,{\cal T}^2[G_{\rm L}]+\frac{\lambda_0^{((N-1)A)}}{12N}{\cal T}[G_{\rm L}]{\cal T}[G_{\rm T}]+\frac{\lambda_0^{((N-1)^2 A+2(N-1)B)}}{24N}\,{\cal T}^2[G_{\rm T}]\nonumber\\
&& 
+\frac{\phi^2}{12N}\big[\lambda_2^{(A+2B)}{\cal T}[G_{\rm L}]+\lambda_2^{((N-1)A)}{\cal T}[G_{\rm T}]\big]
-\frac{\lambda^2_\star\phi^2}{36 N^2}\big[3{\cal S}[G_{\rm L}]+(N-1){\cal S}[G_{\rm L};G_{\rm T};G_{\rm T}]\big]\,,
%+\frac{\lambda_2^{(A+2B)}}{12N}\phi^2{\cal T}[G_{\rm L}]+\frac{\lambda_2^{((N-1)A)}}{12N}\phi^2{\cal T}[G_{\rm T}]
%-\frac{\lambda^2_\star}{12N^2}\phi^2{\cal S}[G_{\rm L}]-\frac{(N-1)\lambda^2_\star}{36N^2}\phi^2{\cal S}[G_{\rm L};G_{\rm T};G_{\rm T}]\,,
\eeq
\end{widetext}
but instead of evaluating it at its stationary point, defined by the solution of Eqs.~(\ref{eq:bml}) and (\ref{eq:bmt}), we evaluate it at the stationary point of the Hartree-Fock effective potential. There are two main reasons to do this here. Since the Hartree-Fock gap equations are equations for a momentum independent self-energy, the possibility rises to draw some conclusions, including renormalization, analytically and also numerical calculations become faster, allowing for a thorough investigation of the model.\footnote{We shall see that, once a physical parametrization of the model is performed, our results in the two-loop and hybrid approximations will not differ much.} The momentum independence of the self-energy also allows us to conveniently work in dimensional regularization. We stress however that what follows can be redone equivalently using a cutoff regularization. We note finally that the need for $\gamma_0(m_\star)$ in the expression (\ref{eq:2loop_effpot}) stems from a proper regularization of the 2PI effective action, as discussed in Ref.~\cite{Marko:2012wc}.  In dimensional regularization, we have\footnote{There was a factor of $1/2$ missing in Eq.~(22) of Ref.~\cite{Marko:2012wc} and as a consequence there should be a factor of 2 in front of the two terms of the last line of Eq.~(31) of that reference.}
\beq\label{Eq:gamma_0}
\gamma_0(m_\star)=\frac{1}{2}\int\frac{d^{d-1}q}{(2\pi)^{d-1}}\Big[\varepsilon_q^\star+2T\ln\left(1-e^{-\varepsilon^\star_q/T}\right)\Big]\,,\quad\ \  
\eeq
with $d=4-2\epsilon.$\\

Although it can be performed explicitly, see below, the discussion of renormalization in the hybrid case is more subtle than in the $\Phi$-derivable case, because two different levels of approximation for the 2PI effective potential are intertwined, each of which comes with its own set of counterterms. First, the Hartree-Fock effective potential, from which the hybrid gap equations are deduced, is obtained from Eq.~(\ref{eq:2loop_effpot}) after removing the setting-sun sum integrals, making the replacements $m_2\to m_0$ and $\lambda_2^{(A,B)}\to\lambda_0^{(A,B)}$, and taking $\lambda_4$ as given by Eq.~(\ref{eq:l4H}). The parameters $m_2$ and $\lambda_2^{(A,B)}$ are taken equal to $m_0$ and $\lambda_0^{(A,B)}$ because, in the Hartree approximation and at $\phi=0$, $\hat M^2=\bar M^2$ and $V=\bar V$. To understand why (\ref{eq:l4H}) is the relevant choice for $\lambda_4$, one has to recompute $\hat V$ in the Hartree approximation, along the lines of Appendix~\ref{app:four-point}. It follows in particular that the gap equations in the hybrid approximation read
\beq\label{eq:bmlh}
\bml&=&m_0^2+\frac{\lambda_0^{(A+2B)}}{6N}\left[\phi^2+{\cal T}[\bar G_{\rm L}]\right]\nonumber\\
&& +\frac{\lambda_0^{((N-1)A)}}{6N}{\cal T}[\bar G_{\rm T}]
\eeq
and
\beq\label{eq:bmth}
\bmt&=&m_0^2+\frac{\lambda_0^{(A)}}{6N}\left[\phi^2+{\cal T}[\bar G_{\rm L}]\right]\nonumber\\
&&+\frac{\lambda_0^{((N-1)A+2B)}}{6N}{\cal T}[\bar G_{\rm T}]\,,
\eeq
which are obtained equivalently from Eqs.~(\ref{eq:bml}) and (\ref{eq:bmt}) by disregarding the momentum dependent pieces and making the replacements $\lambda_2^{(A,B)}\to\lambda_0^{(A,B)}$. Second, since the two-loop 2PI effective potential is evaluated for a different propagator than in the two-loop case,  the bare parameters needed to renormalize the hybrid effective potential need not be the same as those derived in the previous section. In fact, it can be immediately seen that $m_0, m_2, \lambda_0^{(A,B)},\lambda_2^{(A,B)}$ are the same as they are also needed to renormalize the gap and curvature masses at $\phi=0$ which remain unchanged. However, as mentioned above, the field equation receives additional contributions and therefore $\lambda_4$ is changed. Another point of view is that the four-point function $\hat V_{\phi=0}$ is modified as compared to the two-loop $\Phi$-derivable case. After some calculation whose details are gathered in Appendix~\ref{app:hybrid} and upon imposing the renormalization condition $\hat V_{\phi=0,T_\star}=\lambda_\star$, we arrive at
\beq
\label{eq:lambda4_hybrid}
\lambda_4=-2\lambda_\star+2\lambda_{2{\rm l}}^{(A+2B)}-\lambda_0^{(A+2B)},
\eeq
which gives $\lambda_4=-2\lambda_\star+6\lambda_{2{\rm l}}-3\lambda_0$ when $N=1$.

One of the nice features of the hybrid approximation is that, since the self-energies are momentum independent, the divergent part of the various sum integrals involved in the calculation can be determined analytically. In this way, one can check explicitly that the above counterterms renormalize the gap and field equations as well as the effective potential and explicitly finite expressions can be obtained for them. We will now show in detail how to derive the finite gap equations and then sketch the derivation of the finite hybrid effective potential which in turn leads to a finite field equation by differentiation.\\

%%%
\subsection{Explicit renormalization}
Recall first how the renormalization of the gap equation for $\phi=0$ works. We have seen in this case that everything boils down to a single equation, Eq.~(\ref{eq:bmz}), similar to the gap equation in the case $N=1$. After using the value of $m^2_0$, one obtains
\beq\label{eq:toto}
\frac{\bmz-m^2_\star}{\lambda_0^{(NA+2B)}}=\frac{1}{6N}\left[{\cal T}[\bgz]-{\cal T}_\star[G_\star]\right].
\eeq
By using the techniques developed in Ref.~\cite{Marko:2012wc} or by performing an explicit calculation, it is easily seen that the remaining divergence in the right-hand side is nothing but $-(\bmz-m^2_\star){\cal B}_\star[G_\star](0)/6N$. Subtracting this contribution from both sides of the equation and using Eq.~(\ref{eq:l0NA2B}), we end up with
\beq\label{Eq:bmz}
\bmz=m^2_\star+\frac{N+2}{6N}\lambda_\star {\cal T}_{\rm F}[\bgz]\,,
\eeq
where we have introduced the finite combination
\beq\label{eq:tf}
{\cal T}_{\rm F}[G]&=&{\cal T}[G]-{\cal T}_\star[G_\star]+(M^2-m_\star^2){\cal B}_\star[G_\star](0)\,.\qquad
\eeq
In order to generalize these manipulations to the case $\phi\neq 0$, we note that what matters when $\phi=0$ is that the combination of masses appearing in the left-hand side of Eq.~(\ref{eq:toto}) is exactly the same as the combination of tadpoles in the right-hand side and also that the combinations of bare couplings $\lambda_0^{(A)}$ and $\lambda_0^{(B)}$ are precisely the ones given in Eq.~(\ref{eq:l0NA2B}) whose inverse is finite up to a bubble diagram with the appropriate prefactor. If we were able to find linear combinations of the masses $\bml$ and $\bmt$ involving the same linear combinations of the corresponding tadpoles, we could apply the previous procedure twice. Now, if we write the system of gap equations as
\beq
\left(
\begin{array}{c}
\bml\\
\bmt
\end{array}
\right)=\left(
\begin{array}{cc}
a & b\\
c & d
\end{array}
\right)\left(
\begin{array}{c}
{\cal T}[\bgl]\\
{\cal T}[\bgt]
\end{array}
\right)
+\left(
\begin{array}{c}
u\\
v
\end{array}
\right)\,,
\eeq
we see that if $P$ denotes the matrix that diagonalizes the system, we obtain
\beq
P\left(
\begin{array}{c}
\bml\\
\bmt
\end{array}
\right)=\left(
\begin{array}{cc}
\lambda & 0\\
0 & \mu
\end{array}
\right)P\left(
\begin{array}{c}
{\cal T}[\bgl]\\
{\cal T}[\bgt]
\end{array}
\right)
+P\left(
\begin{array}{c}
u\\
v
\end{array}
\right),\quad
\eeq
and thus $P$ provides the sought-after combinations. These are found to be $\bml+(N-1)\bmt$ and $\bml-\bmt$ and we note that the corresponding equations not only involve by construction the same combinations of tadpoles, that is ${\cal T}[\bgl]+(N-1){\cal T}[\bgt]$ and ${\cal T}[\bgl]-{\cal T}[\bgt]$, but also that they involve respectively the combinations $\lambda_0^{(NA+2B)}$ and $\lambda_0^{(2B)}$ which are those whose inverse is finite up to a bubble diagram with the appropriate prefactors, see Eqs.~(\ref{eq:l02B}) and (\ref{eq:l0NA2B}). We can now apply twice the procedure used for the case $\phi=0$ and, after switching back to longitudinal and transverse components, we finally end up with the equations
\beq\label{eq:bmlfinite}
\bml&=&m_\star^2+\frac{\lambda_\star}{2N}\left [\phi^2+{\cal T}_{\rm F}[\bar G_{\rm L}]\right]
%\nonumber\\&&
+\frac{N-1}{6N}\lambda_\star{\cal T}_{\rm F}[\bar G_{\rm T}]\,,\quad\ \  
\eeq
and
\beq\label{eq:bmtfinite}
\bmt&=&m_\star^2+\frac{\lambda_\star}{6N}\left[\phi^2+{\cal T}_{\rm F}[\bar G_{\rm L}]\right]
%\nonumber\\&&
+\frac{N+1}{6N}\lambda_\star{\cal T}_{\rm F}[\bar G_{\rm T}]\,,\quad\ \ 
\eeq
which are both finite. A similar approach has been used in Ref.~\cite{AmelinoCamelia:1996hw} in the case of a theory with two scalar fields, not related to each other by $O(2)$ symmetry. Surprisingly, the author did not use this approach in the case of the $O(N)$ model in Ref.~\cite{AmelinoCamelia:1997dd}. This is probably related to the fact that he was not considering multiple bare couplings as we do here, see also the discussion below.\\

To sketch the renormalization of the two-loop hybrid effective potential, let us consider the case $N=1$ first. The trick is to express the hybrid effective potential in terms of the Hartree-Fock effective potential 
\beq
\gamma_{\rm H}(\phi)&=&\gamma_0(m_\star)+\frac{m_0^2}{2}\phi^2+\frac{\lambda_4^{\rm H}}{24}\phi^4+\frac{\lambda_0}{8}\big[{\cal T}[\bar G]+2\phi^2\big]{\cal T}[\bar G]\nonumber \\
&+&\frac{1}{2}\int_Q^T\big[\ln\bar G^{-1}-\ln G_\star^{-1}+(Q^2+m_0^2)\bar G -1\big],\nonumber\\
\eeq
which we know how to renormalize, see Ref.~\cite{Reinosa:2011ut}. We have
\beq
\gamma(\phi)& = & \gamma_{\rm H}(\phi)+\frac{m^2_2-m^2_0}{2}\phi^2+\frac{\lambda_{2{\rm l}}-\lambda_0}{4}\phi^4\nonumber\\
&& +\frac{\lambda_2-\lambda_0}{4}\phi^2{\cal T}[\bar G]-\frac{\lambda^2_\star}{12}\phi^2{\cal S}[\bar G]\,,
\eeq
where we have used the fact that in the Hartree-Fock approximation $m_2$ is equal to $m_0$, $\lambda_2$ is equal to $\lambda_0$ and $\lambda_4$ is given by Eq.~(\ref{eq:l4H}) instead of Eq.~(\ref{eq:lambda4_hybrid}), so that the difference accounts for the $\phi^4$ term above. Using the expressions for $m_2$ and $m_0$, together with the gap equation at $N=1$, this term cancels and we arrive at
\beq
\gamma(\phi)=\gamma_{\rm H}(\phi)+\frac{\lambda^2_\star}{4}\phi^2{\cal C}[\bar G,G_\star]\,,
\eeq
with
\begin{widetext}
\beq\label{eq:calc}
{\cal C}[\bar G,G_\star] & = & \frac{2}{\lambda^2_\star}\left(\frac{\lambda_{2{\rm l}}}{\lambda_0}-1\right)(\bar M^2-m^2_\star)+\frac{\delta\lambda_{2{\rm nl}}}{\lambda^2_\star}\left[{\cal T}[\bar G]-{\cal T}_\star[G_\star]\right]-\frac{1}{3}\left[{\cal S}[\bar G]-{\cal S}_\star[G_\star]\right]\nonumber\\
& = & {\cal T}_{\rm F}[\bar G]{\cal B}_\star[G_\star](0)-\frac{1}{3}\left[{\cal S}[\bar G]-{\cal S}_\star[G_\star]-(\bar M^2-m_\star^2)\frac{d{\cal S}_\star[G_\star]}{dm_\star^2}\right].
\eeq
\end{widetext}
The second line has been obtained by using the explicit expressions for $\lambda_{2{\rm l}}$ and $\delta\lambda_{2{\rm nl}}$ and shows that the determination of ${\cal C}[\bar G,G_\star]$ relies essentially on the determination of ${\cal S}[\bar G]$. An explicit proof of the finiteness of ${\cal C}[\bar G,G_\star]$ is given in Appendix~\ref{app:hybrid}. This concludes the proof that the hybrid potential is finite in the case $N=1.$ We mention that a finite expression of ${\cal C}[\bar G,G_\star],$ which can be used for the numerical evaluation of the effective potential, was obtained within dimensional regularization in Ref.~\cite{Marko:2012wc}, see Eq.~(B11) there. 

Similar considerations for arbitrary $N$ lead to
\beq
\label{eq:hyb_effpot_N}
\gamma(\phi) &=&\gamma_{\rm H}(\phi)+\frac{\lambda_\star\phi^2}{36N}{\cal C}_N[\bar G_{\rm L},\bar G_{\rm T},G_\star],
\eeq
where
\beq
\nonumber
{\cal C}_N[\bar G_{\rm L},\bar G_{\rm T},G_\star]&=&(N+8){\cal C}[\bar G_{\rm L},G_\star]\\
&&+(N-1)\tilde{\cal C}[\bar G_{\rm L},\bar G_{\rm T},G_\star]\,,
\label{Eq:C_N}
\eeq
with ${\cal C}$ given in Eq.~(\ref{eq:calc}) and
%\begin{widetext}
\beq
&&\tilde{\cal C}[\bar G_{\rm L},\bar G_{\rm T},G_\star]=2{\cal T}_{\rm F}[\bar G_{\rm T}]{\cal B}_\star[G_\star](0)-\frac{1}{3}\bigg[3{\cal S}[\bar G_{\rm L};\bar G_{\rm T};\bar G_{\rm T}]\nonumber\\
&&\qquad-{\cal S}[\bar G_{\rm L}]-2{\cal S}_\star[G_\star]-2(\bmt-m_\star^2)\frac{d{\cal S}_\star[G_\star]}{dm_\star^2}\bigg].\label{Eq:C_tilde}
\eeq
%\end{widetext}
As it was the case for ${\cal C}$, it is possible to show that $\tilde {\cal C}$ is finite and we refer to Appendix~\ref{app:hybrid} for the details.

Thus, it remains to be shown that the Hartree-Fock potential is renormalized for arbitrary $N$. In fact we expect it to be finite up to a temperature and field independent divergent constant. For this reason, we consider instead the subtracted potential $\Delta\gamma(\phi)\equiv\gamma(\phi)-\gamma_\star(0)$. In the case $N=1$, one possibility is to rewrite the effective potential in terms of the combination $\phi^2+{\cal T}[\bar G].$ We complete a square of the form $\propto(\phi^2+{\cal T}[\bar G])^2$ and gather the terms proportional to the bare mass into $m^2_0(\phi^2+{\cal T}[\bar G]),$ then we use the gap equation to write $\phi^2+{\cal T}[\bar G]=2(\bar M^2-m^2_0)/\lambda_0$ and the expression for $1/\lambda_0,$ which can be read off from Eq.~\eqref{eq:l0NA2B}. Performing these steps we end up with the subtracted effective potential
\beq\label{eq:gamma_ren_N1}
\Delta\gamma_{\rm H}(\phi)&=&\frac{\lambda_4^{\rm H}-3\lambda_0}{24}\phi^4
+\frac{1}{2}\big({\cal L}_{\rm F}[\bar G]-\bar M^2{\cal T}_{\rm F}[\bar G]\big)\nonumber\\
&&+\frac{\bar M^4-m^4_\star}{2\lambda_\star}\,,
\eeq
where we have introduced the subtracted logarithmic sum integral
%\begin{widetext}
\beq
{\cal L}_{\rm F}[G]&\equiv&2\big[\gamma_0(m_\star)-\gamma_0^\star(m_\star)\big]+\int_Q^T\big[\ln G^{-1}-\ln G^{-1}_\star\big]\nonumber\\
&-&(M^2-m_\star^2){\cal T}_\star[G_\star]+\frac{1}{2}(M^2-m_\star^2)^2{\cal B}_\star[G_\star](0)\,,\nonumber\\
\eeq
%\end{widetext}
which can be checked to be finite. It remains to be shown that the combination $\lambda_4^{\rm H}-3\lambda_0$ is finite. From Eq.~(\ref{eq:l4H}), we have $\lambda_4^{\rm H}-3\lambda_0=-2\lambda_\star$, which concludes the proof in the one-component case.

The extension to $N\neq 1$ is rendered difficult by the presence of terms of the form ${\cal T}[\bgl] {\cal T}[\bgt]$ which couple longitudinal and transverse components. However, if one expresses the Hartree-Fock potential in terms of the diagonalizing combinations obtained above, namely $\bgl+(N-1)\bgt$ and $\bgl-\bgt$, one checks that such types of coupled terms disappear. Moreover the combinations of bare couplings which come with such decoupled combinations are again precisely those for which we have simple expressions given by Eqs.~(\ref{eq:l02B}) and (\ref{eq:l0NA2B}). We can thus repeat twice the standard procedure for the case $N=1$. After switching back to longitudinal and transverse components, we finally end up with the renormalized expression
\begin{widetext}
\beq\label{Eq:V_hybrid_final}
\nonumber
\Delta\gamma_{\rm H}(\phi)&=& -\frac{\lambda_\star\phi^4}{12N}+\frac{1}{2}({\cal L}_{\rm F}[\bar G_{\rm L}]-\bml{\cal T}_{\rm F}[\bar G_{\rm L}])+\frac{N-1}{2}({\cal L}_{\rm F}[\bar G_{\rm T}]-\bmt{\cal T}_{\rm F}[\bar G_{\rm T}])\\
%\nonumber
&& +\frac{3N}{(N+2)\lambda_\star}\left[\frac{N+1}{4}(\bar M_L^4-m_\star^4)+\frac{3(N-1)}{4}(\bar M_T^4-m_\star^4)-\frac{N-1}{2}(\bml\bmt-m_\star^4)\right].
\eeq
\end{widetext}
In the hybrid approximation we shall not use the field equation, but search for the minimum of the effective potential \eqref{Eq:V_hybrid_final}, as explained in Sec.~\ref{sec:numerics}, nevertheless, for completeness, we give its renormalized form in Appendix~\ref{app:hybrid}.

%%%
\subsection{Comparison to other approaches}
To close this section, let us compare our renormalization procedure to other approaches followed in the literature. Since most of these approaches concern the Hartree-Fock approximation, we focus on the latter for which we have given the renormalized effective potential in Eq.~(\ref{Eq:V_hybrid_final}) and the renormalized gap  equations in (\ref{eq:bmlfinite})-(\ref{eq:bmtfinite}). The finite field equation can be obtained by plugging Eq.~(\ref{eq:bmlh}) into the Hartree-Fock bare field equation to yield
\beq\label{eq:field_2}
0=\bml+\frac{\lambda^{\rm H}_4-\lambda_0^{(A+2B)}}{6N}\bar\phi^2=\bml-\frac{\lambda_\star}{3N}\bar\phi^2\,,\ \ 
\eeq
where we have also used Eq.~(\ref{eq:l4H}).\\ 

The renormalization of the Hartree-Fock approximation was investigated for instance in Ref.~\cite{Lenaghan:1999si} where two different regularization schemes, cutoff and dimensional regularization, were used together with the corresponding ``renormalization'' schemes, named respectively ``cutoff scheme'' (CO) and ``counterterm scheme'' (CT) and leading surprisingly to different results. In fact the CO scheme is not really a renormalization scheme since the authors explain that there is no way to send $\Lambda$ to infinity and the equations need to be considered at finite $\Lambda$, $\Lambda$ being an additional parameter of the model. The drawback of such an approach is that certain obstructions appear in parameter space, in particular in the chiral limit. In contrast, the CT scheme removes the divergences and the continuum limit can be considered, with no obstruction in the chiral limit. This seems contradictory since one could expect that physical results should not depend on the regularization method used. Moreover, the CT scheme was not given a real justification in Ref.~\cite{Lenaghan:1999si} and it was not clear how to generalize it to higher order truncations. The renormalization that we use in this work clarifies these issues. As we now explain, it gives a justification to the CT scheme of Ref.~\cite{Lenaghan:1999si}, it is generalizable to an arbitrary level of truncation and it allows us to modify the CO scheme in such a way that it becomes formally identical to the CT scheme. In particular it presents no obstructions in the chiral limit. 

If we have a closer look at  Eqs.~(\ref{eq:bmlfinite}) and (\ref{eq:bmtfinite}) for instance, we notice that, except for the fact that the subtractions are made at a finite temperature $T_\star$, our renormalized gap equations have structurally the same form as those of the CT scheme of Ref.~\cite{Lenaghan:1999si}. We thus see that one way to justify this scheme is to admit the need for multiple bare parameters\footnote{As it is explained in Ref.~\cite{Reinosa:2011ut}, the need for multiply defined bare parameters is a truncation artifact and, the consistency conditions are such that, if one increases the order of the truncation the differences between the various bare parameters, should become smaller and smaller, at least formally.} which need to be fixed by appropriate renormalization conditions, supplemented by consistency conditions. Unlike what is stated in Ref.~\cite{Lenaghan:1999si}, our interpretation shows that the CT scheme does not involve temperature dependent counterterms since the counterterms depend only on the renormalization scale $T_\star$ but not on the self-consistent mass $\bar M^2$. Moreover, these considerations are sufficiently general to be extendable to higher order approximations or to apply to any regularization, with similar results in the continuum limit. In particular, we can define a CO scheme for which the renormalized equations are (\ref{eq:bmlfinite}) and (\ref{eq:bmtfinite}) with integrals cut off at some scale $\Lambda$. In this scheme the cutoff $\Lambda$ can be sent to infinity (as mentioned above, this is a peculiarity of lower order approximations) and no obstructions appear in the chiral limit. In fact the problems with the CO scheme in Ref.~\cite{Lenaghan:1999si} can all be identified with the use of one single bare coupling, instead of multiple ones as we propose here. To illustrate this, let us revisit one of the obstructions raised in Ref.~\cite{Lenaghan:1999si} and see how it is lifted within our approach. In the CO scheme of Ref.~\cite{Lenaghan:1999si}, the gap equations at finite $\Lambda$ are written using a single bare coupling. This amounts to replacing $\lambda_0^{(A)}$ and $\lambda_0^{(B)}$ by $\lambda_0$ in the bare gap equations (\ref{eq:bmlh}) and (\ref{eq:bmth}). Similarly the bare field equation is written with the same coupling $\lambda_0$ everywhere and reads in the Hartree-Fock approximation:
\beq\label{eq:field_3}
0 & =  & m_0^2+\frac{\lambda_0}{6N}\bar\phi^2+\frac{\lambda_0}{2N}{\cal T}[\bgl]+\frac{(N-1)\lambda_0}{6N}{\cal T}[\bgt]\nonumber\\
& = & \bml-\frac{\lambda_0}{3N}\bar\phi^2\,, 
\eeq
to be compared to Eq.~(\ref{eq:field_2}). Writing the difference of the two gap equations at $T=0$, setting the pion mass to zero and using the field equation (\ref{eq:field_3}), one arrives then at
\beq
0=\int_{Q<\Lambda}^{T=0}\frac{1}{Q^2+m^2_\sigma}-\int_{Q<\Lambda}^{T=0}\frac{1}{Q^2}\,,
\eeq
whose solutions are either $m_\sigma=0$ or $\Lambda=0$, both absurd. This is the conclusion reached in Ref.~\cite{Lenaghan:1999si}. In contrast, within our scheme, if we subtract the renormalized gap equations (\ref{eq:bmlfinite})-(\ref{eq:bmtfinite}) and use renormalized field equation (\ref{eq:field_2}), we obtain
\beq\label{eq:CO2}
0 & = & \int^{T=0}_{Q<\Lambda}\frac{1}{Q^2+m^2_\sigma}-\int^{T=0}_{Q<\Lambda}\frac{1}{Q^2}\nonumber\\
 & & +\,m^2_\sigma\int_{Q_\star<\Lambda}^{T_\star}\frac{1}{(Q_\star^2+m^2_\star)^2}\,,
\eeq
which admits a nonzero solution\footnote{As a function of $m^2_\sigma$, the right-hand side of the equation starts at $0$ when $m^2_\sigma=0$, and decreases first before growing linearly as $m^2_\sigma\to\infty$.} for $m_\sigma$, pretty insensitive to the large values of $\Lambda$ because Eq.~(\ref{eq:CO2}) is renormalized.\\

Our approach differs also from that used by Amelino-Camelia and Pi in Refs.~\cite{AmelinoCamelia:1992nc,AmelinoCamelia:1997dd} where only one bare coupling was used. If we were to use only one bare coupling, the first term of Eq.~(\ref{eq:gamma_ren_N1}) would be $-(\lambda_0/12)\phi^4$ in place of $-(\lambda_\star/12)\phi^4$. According to Amelino-Camelia this term does not spoil the renormalizability because $\lambda_0$, albeit being a bare parameter, approaches $0^-$ as $\Lambda\to\infty$. However, as already discussed above, the possibility to send the cutoff to infinity is a peculiarity of the lowest order approximations, not shared by higher order ones where we expect physical quantities not to be defined above the Landau scale. It is thus more satisfactory to implement a renormalization scheme in which the results are already pretty much insensitive to the cutoff below the Landau scale. This is achieved by our scheme if the Landau scale is not too close to the physical scales because our results show a ``plateau'' behavior below the Landau scale, whereas in the scheme by Amelino-Camelia there remains a logarithmic sensitivity from the term $-(\lambda_0/12)\phi^4$. One could argue that the existence of a plateau is related to the existence of a continuum limit, a notion that does not make sense at higher orders of approximation. Still, as already mentioned above, if the parameters are such that the Landau scale is much larger than the relevant physical scales, there is an intermediate regime where this notion can be considered in a somewhat generalized acceptation: quantities renormalized according to our scheme will still show a plateau behavior for values of the cutoff below the Landau scale. These considerations can be made more quantitative by using specific examples and will be presented elsewhere \cite{WiP}. In our present two-loop approximation we will study the cutoff dependence of some physical quantities for different values of the parameters, that is different values of the Landau pole (see Figure~\ref{Fig:CO_dep}). \\

Let us finally mention that certain works disregard renormalization by arguing that one is only interested in thermal effects and thus that ``vacuum'' fluctuations can be neglected, see for instance Ref.~\cite{Petropoulos:1998gt}. It is worth mentioning however that, in a self-consistent context such as the 2PI formalism, the masses or self-energies that enter these vacuum fluctuations depend on the temperature. Neglecting them is then not completely justified and can lead to neglecting an important piece of the thermal contribution. This can be tested by using the exact limits of certain models/theories such as the limit of a large number of flavors in QED/QCD, see Ref.~\cite{Blaizot:2005wr}.

%%%%%
\section{Numerical method}\label{sec:numerics}

Before discussing our results in the next section, let us give a brief overview of the numerical methods that we used to solve the equations and compute various quantities of interest.

 In the two-loop case we take advantage of the fact that all momentum dependent sum integrals are convolutions and compute them by means of discrete fast Fourier transform algorithms (DST and DCT as described in Refs.~\cite{Borsanyi:2008ar,Marko:2012wc}) using a 3D cutoff $\Lambda.$  We exploit the rotation symmetry of the propagators to reduce our discretization to a two-dimensional $N_\tau\times N_s$ lattice containing $N_\tau-1$ positive Matsubara frequencies in addition to the static mode $\omega_n=0$ and $N_s$ moduli of the 3D momentum, the smallest available being the lattice spacing in momentum space $\Delta k=\Lambda/N_s.$ Moreover, since the leading asymptotic behavior of $\bar G(Q)$ is exactly $1/Q^2$ in the approximation at hand, we can increase the rate of convergence of the Matsubara sums and of the convolutions by subtracting first the leading (free-type) asymptotic behavior of the various summands/integrands.  These subtracted sum integrals involve free-type propagators, as it is also the case for all the sum integrals encountered in the hybrid approximation, and therefore can be computed almost exactly. In practice this means that the Matsubara sum is performed exactly and the momentum integral is computed numerically using accurate adaptive integration routines of the GNU Scientific Library (GSL) \cite{gsl}. For more details on the numerical aspects, we refer to our previous work \cite{Marko:2012wc} and adopt the notations used in its Sec.~V. In the remainder of this section, we describe some of the most important aspects, in particular the new features that appear in the case $N\neq 1$.

%%%

\subsection{Increasing the rate of convergence of the sum integrals}

After using the expressions for $m_0^2,$ $m_2^2$ and $\delta\lambda_{\rm 2nl}^{(A,B)},$ which can be read off from Eqs.~\eqref{Eq:m0_explicit}, \eqref{Eq:m2_explicit}, \eqref{eq:dl2Bnl} and \eqref{eq:dl2NAp2Bnl}, it is straightforward to apply the procedure described in Sec.~V.B of Ref.~\cite{Marko:2012wc} to render the longitudinal gap equation \eqref{eq:bml}, the gap equation at $\phi=0$ \eqref{eq:bmz}, and the expression of the curvature at $\phi=0$ \eqref{eq:curvature} in a form suitable for numerical computations, because they contain the same types of sum integrals as those in Ref.~\cite{Marko:2012wc}. This is true also for the subtracted effective potential, defined using Eq.~\eqref{eq:2loop_effpot} as $\Delta\gamma(\phi)=\gamma(\phi)-\gamma_\star(0),$ when it is written using the expression $\delta\gamma/\delta\phi$ appearing in the field equation\footnote{In the presence of an external field $h,$ $\delta\gamma/\delta\phi$ is the expression appearing on the right-hand side of Eq.~\eqref{eq:field}, but multiplied by $\phi$ and with $h\phi$ subtracted from it.} as
\begin{widetext}
\beq
\nonumber
\Delta\gamma(\phi)&=&N\big(\gamma_0(m_\star,\Lambda)-\gamma_0^\star(m_\star,\Lambda)\big)
+\sum_{i={\rm T,L}}\frac{c_i}{2}\int_Q^T\big[\ln \bar G_i^{-1}(Q)-\ln G_\star^{-1}(Q)-(\bar M_i^2(Q)-m_\star^2)\bar G_i(Q)\big]
\\
\nonumber
&&+\frac{1}{2}\phi\frac{\delta\gamma}{\delta\phi}-\frac{\lambda_4}{24N}\phi^4-\frac{h\phi}{2}
+\frac{\lambda_0^{(A)}}{24N}\Big[{\cal T}[\delta \bar G_{\rm L}]+(N-1){\cal T}[\delta \bar G_{\rm T}]+N\delta{\cal T}[G_\star]\Big]^2\\
&&+\frac{\lambda_0^{(B)}}{12N}\left[\big({\cal T}[\delta \bar G_{\rm L}]+\delta{\cal T}[G_\star]\big)^2+(N-1)\big({\cal T}[\delta \bar G_{\rm T}]+\delta{\cal T}[G_\star]\big)^2\right],
\eeq
\end{widetext}
where $c_{\rm L}=1,$ $c_{\rm T}=N-1,$ $\gamma_0^\star(m_\star,\Lambda)$ is the integral in Eq.~\eqref{Eq:gamma_0} calculated with a cutoff $\Lambda$ and at a temperature $T_\star$ and we used the shorthand notations $\delta \bar G_{\rm L/T}=\bar G_{\rm L/T}-G_\star$ and $\delta{\cal T}[G_\star]={\cal T}[G_\star]-{\cal T}_\star[G_\star].$  The integrals are evaluated as shown in Eq.~(131) of Ref.~\cite{Marko:2012wc}.  However, for the transverse gap equation \eqref{eq:bmt} and the field equation itself \eqref{eq:field}, we need to compute two sum integrals which were not encountered in our previous work. The first is the bubble sum integral of Eq.~\eqref{eq:bmt}, which is rewritten as
\beq
\nonumber
{\cal B}[\bar G_{\rm L};\bar G_{\rm T}](K)&=&{\cal B}[G_\star](K)+\int_Q^T\bar G_{\rm L}(Q)\delta\bar G_{\rm T}(K-Q)\\
&&+\int_Q^T\delta\bar G_{\rm L}(Q)G_\star(K-Q),
\label{Eq:new_bubble}
\eeq
where $\delta \bar G_{\rm L/T}$ decrease faster in the UV than $\bar G_{\rm L/T},$ hence reducing the error of the corresponding sum integrals, as compared to that of the sum integral ${\cal B}[\bar G_{\rm L};\bar G_{\rm T}]$, while the first term involves only the free-type propagator $G_\star$ and can be computed almost exactly. The discretized form of Eq.~\eqref{Eq:new_bubble} used in the numerics can be easily given in terms of the discrete version of the convolution defined in Eq.~(114) of Ref.~\cite{Marko:2012wc}.  The second new sum integral is decomposed as
\beq
\nonumber
{\cal S}[\bar G_{\rm L};\bar G_{\rm T};\bar G_T]&=&\int_Q^T\bar G_{\rm L}(Q)({\cal B}[\bar G_{\rm T}](Q)-{\cal B}[G_\star](Q))\\
&&+\int_Q^T\delta\bar G_{\rm L}(Q){\cal B}[G_\star](Q)+{\cal S}[G_\star].
\eeq
The third term and the bubble ${\cal B}[G_\star](Q)$ in the second term can be computed almost exactly. The summand in the second term decreases faster than the original one,  $\bar G_{\rm L}(Q) {\cal B}[\bar G_{\rm T}](Q),$ as it is the case with the difference of bubbles in the first term, which can be rewritten as a convolution using Eq.~(121) of Ref.~\cite{Marko:2012wc}. Then, the discretized form of ${\cal S}[\bar G_{\rm L};\bar G_{\rm T};\bar G_T]$ can be readily written using the discrete version of the convolution and of the local sum integral defined in Eqs.~(114) and (115) of Ref.~\cite{Marko:2012wc}. 

%%%
\subsection{On the solution of the equations}

In the two-loop approximation the solution of the gap equations \eqref{eq:bml} and \eqref{eq:bmt} either at fixed $\phi$ or together with the field equation \eqref{eq:field} is obtained iteratively. In both cases the coupling counterterms $\lambda_0^{(A,B)}$, $\lambda_{2\rm l}^{(A,B)}$ and $\lambda_4$ are evaluated first using accelerated Matsubara sums, as explained in Appendix~C of Ref.~\cite{Marko:2012wc}. Then, the $T$-dependent integrals which do not depend on the solution of the equations are evaluated using adaptive numerical integration routines. The quantities determined up to this point are unchanged during the iterative process. The process used to solve the coupled equations \eqref{eq:bml}, \eqref{eq:bmt} and \eqref{eq:field} at $h\ne0$ is similar to that used in Ref.~\cite{Marko:2012wc}. At a given $T$ both propagators are initialized with $G_\star.$ The iteration starts with the evaluation, using the most recent $\bar G_{\rm L/T},$ of the local-type sum integrals in the field equation, which is easily solved for it is cubic in $\bar\phi.$ Using the obtained value of $\bar\phi,$ the propagators are updated sequentially, starting with $\bar G_{\rm L}.$ First, the self-energy $\bar M_{\rm L}^2(i\omega_n, k)$ is evaluated by computing the required sum integrals with the most recent propagators (due to the sequential update of the propagators there is no need to recalculate all the local-type sum integrals). Then, the updated propagator is
\beq
\bar G_{\rm L}(i\omega_n, k)=[\omega_n^2+k^2+\alpha\bar M_{\rm L}^2+(1-\alpha)\bar M_{\rm L,old}^2]^{-1},\ \ 
\eeq
where ``old'' refers to the propagator of the previous iteration, which has to be stored. The updated $\bar G_{\rm L}$ is then used to update $\bar G_{\rm T}$ in an analogous way, using the same $\alpha\in(0,1]$ parameter, which controls the speed of convergence of the iterative process. For large $\lambda_\star$ one needs $\alpha<1$ for the iteration procedure to converge at all, however, for small couplings the fastest convergence is achieved with $\alpha=1.$ Besides $\bar\phi$, the value of the propagators at the lowest available frequency and momentum is also monitored. The iteration stops when the relative change of all these quantities from one iteration to the next is smaller than the desired accuracy (usually a relative change smaller than $10^{-7}$ was required).\\

In the hybrid approximation the gap equations \eqref{eq:bmlfinite} and \eqref{eq:bmtfinite} are momentum independent and, therefore, much easier to solve compared to the full two-loop case. However, the field equation is complicated due to the fact that the propagators do not fulfill the stationarity conditions. For this reason, we evaluate instead the effective potential \eqref{eq:hyb_effpot_N} and search for its minimum. During this process the vacuum parts of the sum integrals can be calculated analytically, while the explicitly temperature dependent parts can be computed almost exactly using adaptive numerical integration. Note that one can avoid the determination of $\bar M_{\rm L/T}$ as a solution of two coupled equations. In the next section, we will see in Eq.~\eqref{eq:bml_w_bmt} that it is possible to explicitly express $\bar M_{\rm L}$ in terms of $\bar M_{\rm T}$. Plugging this expression into Eq.~\eqref{eq:bmtfinite} yields a one-dimensional equation for $\bar M_{\rm T},$ to be solved for any $\phi$. Then, $\bar M_{\rm T}, \bar M_{\rm L}$ at the minimum of the potential are easily obtained with a numerical minimum finder routine, which chooses values of $\phi$ and checks the value of the effective potential \eqref{eq:hyb_effpot_N} evaluated with the solution $\bar M_{\rm T}$ of the one-dimensional gap equation and $\bar M_{\rm L}$ determined from Eq.~\eqref{eq:bml_w_bmt}.

%%%
\subsection{Determination of the (pseudo-) critical temperature and zero temperature quantities}
In the chiral limit the critical temperature $T_{\rm c}$ is the value at which the curvature of the potential vanishes at $\phi=0.$ Since the latter is the same in both approximations considered in this work, the corresponding $T_{\rm c}$ is also the same. Moreover since the gap equation \eqref{eq:bmz} yields a momentum independent solution, the curvature at vanishing field, and therefore  $T_{\rm c},$ can be evaluated almost exactly, using adaptive integration routines. Using Eqs.~(\ref{Eq:m2_explicit}), (\ref{eq:l0NA2B}) and (\ref{eq:l2lNA2B}) in Eq.~\eqref{eq:curvature} one can even obtain an explicitly finite equation for the curvature at vanishing field in terms of the the gap mass given by Eq.~\eqref{Eq:bmz} and ${\cal C}[\bar G_{\phi=0},G_\star],$ defined in Eq.~\eqref{eq:calc}:
\beq\label{Eq:curvature_finite}
\hat M^2_{\phi=0}=\bar M_{\phi=0}^2+\frac{N+2}{6N^2}\lambda_\star^2{\cal C}[\bar G_{\phi=0},G_\star].
\eeq
The critical temperature $T_{\rm c}$ is then obtained from the previous expression through the relation $\hat M_{\phi=0,T_{\rm c}}=0$. It is also convenient to define a temperature $\bar T_{\rm c}$ from the vanishing of the gap mass at $\phi=0$: $\bar M_{\phi=0,\bar T_{\rm c}}=0$. The temperature $\bar T_{\rm c}$ is the same in both approximations and can be given analytically, since by setting $\bar M_{\phi=0}$ to zero in Eq.~\eqref{Eq:bmz} and introducing 
\beq\label{Eq:Cstar}
%\nonumber
%C_\star&=&m_\star^2+\frac{N+2}{6N}\lambda_\star\Big[\frac{m^2_\star}{16\pi^2}-{\cal T}_\star^{(1)}[G_\star]\\
%&&-m_\star^2{\cal B}_\star^{(1)}[G_\star](0)\Big],
C_\star=m_\star^2+\frac{N+2}{6N}\lambda_\star {\cal T}_{{\rm F},T=0}[G_0],
\eeq
with $G_0(Q)\equiv 1/Q^2$, as in the one-component case in Ref.~\cite{Reinosa:2011ut}, one obtains $\bar T_{\rm c}=[-72 N C_\star/((N+2)\lambda_\star)]^{1/2},$ if the parameters are such that $C_\star\le0,$ otherwise it is not defined. We note that because the gap and curvature masses admit a continuum limit, so do the critical temperatures. Of course these continuum values are not directly connected with the critical temperatures of the systems the model could describe at low energies, because these are nonuniversal quantities which depend on the microscopic details of the particular system under study. To obtain them, one should rather envisage a first principle calculation or include sufficiently enough nonrenormalizable operators in the model.\\

In the physical case, i.e. at nonzero $h,$ the pseudocritical temperature $T_{\rm pc}$ is defined through the inflection point of the $\bar\phi(T)$ curve, which is determined in both approximations with the same algorithm. This takes into account that since in the two-loop case the computation is time demanding, it becomes worthwhile to determine the inflection point by running the code at the least possible number of temperature values without giving up the accuracy requirements. As a first step of the algorithm we compute $\bar\phi$ at five equidistant temperature values between $T_{\rm c}$ and ${\rm min}(3\bar\phi(T_{\rm c}),5T_\star)$ (this proved always larger than $T_{\rm c}$), where $T_{\rm c}$ is the critical temperature corresponding to the actual value of the parameters $m^2_\star/T^2_\star$ and $\lambda_\star,$ but $h=0.$ Then, from this set of points we compute numerically the first and second derivatives, using the highest possible order of finite difference formulas for central or one-sided approximations \cite{fdf} which can be reached at a certain value of the temperature, given the finite number of points we have. Using the information that $d\bar\phi/dT$ has a minimum at $T=T_{\rm pc}$ and $d^2\bar\phi/dT^2$ changes sign as it goes through $T=T_{\rm pc},$ where it vanishes, we can determine from our five points the two values of temperatures $T_<$ and $T_>$ which enclose the inflection point ($T_<<T_{\rm pc}<T_>$). Next, a rough estimate for the pseudocritical temperature, $T_{\rm est},$ is obtained from $T_<$ and $T_>$ through a linear interpolation. Finally, we compute $\bar\phi$ at three more temperatures: $(T_{\rm est}+T_<)/2,$ $T_{\rm est},$ $(T_{\rm est}+T_>)/2$ and by fitting the function $f(T)=a+b\arctan{c(\pi-d T)}$ to the values of $\bar\phi$ available at these temperature values and at $T_<$ and $T_>,$ we obtain our best estimate for the abscissa of the inflection point: $T_{\rm pc}=\pi/d.$\\

The determination of the $T=0$ quantities required for the parametrization of the model or for the computation of the pressure, see below,  is different in the two approximations that we consider. In the hybrid case the vacuum parts of the integrals can be evaluated analytically, rendering the $T=0$ value of $\hat M_{\rm L/T}$ and $\bar\phi$ or even the effective potential easily accessible. On the contrary, in the two-loop approximation it is impossible to explicitly reach $T=0$ due to the use of a finite number of Matsubara frequencies, since the number of needed frequencies is inversely proportional to the temperature. However, this shortcoming of the numerical method is overcome with an extrapolation procedure which uses the low temperature data obtained by increasing the value of $N_\tau$ to an appropriate value (see the caption of Figure~\ref{Fig:masses} for an explicit example). The effective potential at $T=0$ is obtained by fitting to the low-$T$ values a functional form based on the temperature dependence of the ideal gas pressure, $g(T)=a-bT^{5/2}\exp(-c/T).$ To obtain $\hat M_{\rm L/T}$ and $\bar\phi$ at $T=0$ we use a fitting function $j(T)=a-b\exp(-c/\sqrt{T}),$ which has a purely empirical motivation.

%%%
\subsection{Characteristic curves \label{ss:curves}}
In preparation for the discussion of the results in Sec.~\ref{sec:results}, it is convenient to define certain characteristic curves in the parameter space $(m^2_\star/T^2_\star,\lambda_\star)$.

A first class of curves that we use is made of the iso-$\Lambda_{\rm p}$ curves which allow us to determine a region where the Landau scale is large enough.\footnote{We have treated the hybrid approximation using dimensional regularization and taking the continuum limit after proper renormalization of the equations. We could have proceeded equivalently using a 3D cutoff as in the two-loop case. The iso-$\Lambda_{\rm p}$ curves need to be understood in this context.} We note that, for $N=1,$ Eq.~\eqref{Eq:Lp} goes over into Eq.~(48) of Ref.~\cite{Marko:2012wc},\footnote{Note that there is a factor of 1/2 missing in front of the integral.} which means that the scale of the Landau pole obtained in the $N=1$ case for some value of the coupling is obtained in the $N=4$ case at twice that value. The value of the Landau pole corresponding to a given $\lambda_\star$ can be accurately estimated for $m_\star\ll \Lambda_{\rm p}$ using the formula 
\beq\label{Eq:Lp_approx}
\Lambda_{\rm p}^{\rm est}\approx\frac{m_\star}{2}\exp\left[\frac{48\pi^2N}{(N+2)\lambda_\star}+1-8\pi^2{\cal B}_{\star,\Lambda=\infty}^{(1)}[G_\star](0)\right],
\nonumber\\
\eeq
obtained by replacing in Eq.~\eqref{Eq:Lp} the ``thermal'' part of the bubble integral ${\cal B}_{\star,\Lambda_{\rm p}}^{(1)}[G_\star](0)$ with  ${\cal B}_{\star,\Lambda=\infty}^{(1)}[G_\star](0)$ (we use the notations of Ref.~\cite{Marko:2012wc}).

We also use the $\bar T_{\rm c}=0$ curve, whose equation $\bar\lambda_\star^{\rm c}(m_\star/T_\star)$ can be simply obtained using Eq.~\eqref{Eq:Cstar} from the relation $C_\star=0,$ as\footnote{A simpler expression,  $\bar\lambda_\star^{\rm c}(m_\star/T_\star)\approx(72 N m_\star^2/T_\star^2)/[(N+2)(1-3m_\star/(2\pi T_\star))],$ is obtained using high-temperature expansion, which is reliable for $m_\star/T_\star\lesssim 1$ and sufficient for our purposes.}
\beq\label{Eq:blc_line}
%\bar\lambda_\star^{\rm c}\left(\frac{m_\star}{T_\star}\right)=\frac{6N}{N+2}
%\left[\frac{{\cal T}^{(1)}_\star[G_\star]}{m_\star^2}+{\cal B}_\star^{(1)}[G_\star](0)-\frac{1}{16\pi^2}\right]^{-1}.\nonumber\\
\bar\lambda_\star^{\rm c}\left(\frac{m_\star}{T_\star}\right)=-\frac{6N}{N+2}m^2_\star{\cal T}_{{\rm F},T=0}^{-1}[G_0].
\eeq
This curve can be seen in Figure~\ref{Fig:chiralParams}. For points which are above (below) it $C_\star<0$ ($C_\star>0$). We shall need similarly the curve $T_{\rm c}=0$ whose equation $\lambda_\star^{\rm c}(m_\star/T_\star)$ is obtained implicitly from $\hat M_{\phi=0,T_{\rm c}=0}=0,$
with the renormalized curvature mass given in Eq.~(\ref{Eq:curvature_finite}), that is
\beq\label{Eq:lc_line}
0=\bar M_{\phi=0,T_{\rm c}=0}^2+\frac{N+2}{6N^2}\lambda_\star^2{\cal C}_{T_{\rm c}=0}[\bar G_{\phi=0,T_{\rm c}=0},G_\star].\ \ 
\eeq 
We note that the solution of the gap equation at vanishing temperature (and for $N=1$ even at arbitrary value of the field) can be obtained in closed form in terms of the two real branches of the Lambert function ${\cal W}$. In Appendix~\ref{app:hybrid} we provide the solution at vanishing field and temperature, which can be used in \eqref{Eq:lc_line} to obtain $\lambda_\star^{\rm c}(m_\star/T_\star)$ numerically.

The last characteristic curve is needed in the hybrid case. In the Hartree-Fock approximation applied to the one-component case it was already observed in Ref.~\cite{Reinosa:2011ut} that there is a temperature dependent critical value of the field, $\phi_{\rm c}(T),$ satisfying $\phi_{\rm c}(T)<\bar\phi(T)$, such that for smaller values of the field the gap equation does not admit a physical solution. We investigate now the existence of such a curve in the hybrid approximation and its location with respect to $\bar\phi(T)$. Subtracting three times Eq.~\eqref{eq:bmtfinite} from Eq.~\eqref{eq:bmlfinite} one can express $\bar M_{\rm L}$ in terms of $\bar M_{\rm T}$ as 
\beq
\label{eq:bml_w_bmt}
\bar M_{\rm L}^2=3\bar M_{\rm T}^2-2m_\star^2-\frac{N+2}{3N}\lambda_\star{\cal T}_{\rm F}[\bar G_{\rm T}].
\eeq
Expressing ${\cal T}_{\rm F}[\bar G_{\rm T}]$ from the relation above and plugging it in Eq.~\eqref{eq:bmtfinite}, one obtains
\beq
\nonumber
(1-N)\bar M_{\rm T}^2&=&2m_\star^2-(N+1)\bar M_{\rm L}^2\\
&&+\frac{\lambda_\star}{3 N}(N+2)\big(\phi^2+{\cal T}_{\rm F}[\bar G_{\rm L}]\big).
\label{Eq:gap-T-F_bis}
\eeq
We define $\phi_{\rm c}(T)$ as the value of the field for which $\bar M_{\rm T}(T)=0.$ Then, from Eq.~\eqref{Eq:gap-T-F_bis} one obtains
\beq
\nonumber
\phi^2_{\rm c}(T)=\frac{3N(N+1)}{(N+2)\lambda_\star}\left[\bar M^2_{\rm L,c}(T)-\frac{2 m_\star^2}{N+1}\right]-{\cal T}_{\rm F}[\bar G_{\rm L, c}],\\
\label{Eq:Phi_c_T}
\eeq
where $\bar M^2_{\rm L,c}(T)=\lambda_\star(N+2)(\bar T_{\rm c}^2-T^2)/(36N)$ is obtained from Eq.~\eqref{eq:bml_w_bmt}. Note that, by definition of $\bar T_{\rm c}$, $\bar M_{\rm T}(\bar T_{\rm c})$ vanishes at $\phi=0$. It follows that %at $T=\bar T_{\rm c}$ and $$one has $\bar M_{\rm L,c}(\bar T_{\rm c})=0$ according to the definition of $\bar T_{\rm c}$ given below Eq.~\eqref{Eq:Cstar}, and therefore 
$\bar\phi_{\rm c}(\bar T_{\rm c})=0.$ The existence of the $\bar\phi_{\rm c}(T)$ line depends, of course, on the values of the parameters. It is an interesting question what happens at zero temperature 
%, where one has $\bar M_{\rm L,c}^2(0)=-2 C_\star$ and
%\beq
%\nonumber
%\phi^2_{{\rm c}}(0)&=&\bar M^2_{\rm L,c}(0)\bigg[\frac{3N}{\lambda_\star}-{\cal B}^{(1)}_\star[G_\star](0)\\
%&&+\frac{1}{16\pi^2}\bigg(1-\ln\frac{\bar M^2_{\rm L,c}(0)}{m_\star^2}\bigg)\bigg],
%\label{Eq:Phi_c_T0}
%\eeq
because there could exist a $\bar\phi_{\rm c,0}$ curve in the $(m_\star^2/T_\star^2,\lambda_\star)$ parameter plane along which $\phi_{{\rm c}}(0)=\bar\phi(0)\ne0,$ and which delimits a parameter region where the model cannot be solved at $T=0$ in the hybrid approximation. This is actually the case to the left of the $\bar\phi_{\rm c,0}$ line in the parameter region shown in Figure~\ref{Fig:chiralParams}. We have not seen any trace of a $\bar\phi_{\rm c,0}$ curve in the two-loop approximation.

We mention that we could have alternatively defined $\phi_{\rm c}(T)$ as the value at which $\bar M_{\rm L}(T)$ vanishes. In this case from Eq.~\eqref{Eq:gap-T-F_bis} we have $\bar M^2_{\rm T,c}(T)=-(N+2)\lambda_\star/(3(N-1))\big[\bar\phi_{\rm c}^2(T)+(T^2-\bar T_{\rm c}^2)/12\big],$ which is only positive if for $0\leq T\leq\bar T_{\rm c}$ one has $\phi^2\leq(\bar T_{\rm c}^2-T^2)/12.$ This means that one needs $C_\star\leq 0$. Then similarly as in Appendix~\ref{app:hybrid} (one only has to change $b_\star$ given in Eq.~\eqref{Eq:bmzz} to $b_\star-1/2$) the $T=0$ solution of Eq.~\eqref{eq:bml_w_bmt} at $\bar M_{\rm L}=0$ given in terms of the Lambert function is bigger than $2\Lambda_{\rm p}^{\rm est}\exp[24\pi^2N/((N+2)\lambda_\star)-1].$ The size of this solution matches the size of the $\bar M_{\rm T,c}(0)$ expressed from Eq.~\eqref{Eq:gap-T-F_bis} only for very large $\lambda_\star,$ when the scale of the Landau pole is small. This is outside the region of the parameter space we would like to investigate. It is easy to see, using that ${\cal T}_{\rm F}[G_{\rm T}]$ increases with $T$ at fixed $M_{\rm T},$ that for $0<T<\bar T_{\rm c}$ Eq.~\eqref{eq:bml_w_bmt} admits only a large scale solution, because  the right-hand side of  Eq.~\eqref{eq:bml_w_bmt} at $M_{\rm T}=0$ is negative in this temperature range and vanishes only at $T=\bar T_{\rm c}$. In conclusion, in the region allowed by the definition of $\phi_{\rm c}(T)$ based on the vanishing of $\bar M_{\rm T},$ it turned out that $\bar M_{\rm L}$ is always positive when $\bar M_{\rm T}$ is nonvanishing. 
%However, we have checked numerically that the region that this would exclude in parameter space is included in the region that we have excluded using the definition in terms of $\bar M_{\rm T}$. In fact,
As a last remark we note that in the hybrid approximation for the one-component case, where the only possible definition for $\phi_{\rm c}(T)$ is the one in terms of $\bar M_{\rm L}$, we can prove that the curve $\bar\phi_{\rm c,0}$ does not exist, because, as discussed in Appendix~\ref{app:hybrid}, $\phi_{{\rm c}}(0)=\bar\phi(0)$ cannot happen for $\bar\phi(0)>0,$ that is for parameters for which the model is in its broken phase at $T=0.$

%%%%%
\section{Results}\label{sec:results}
In this section we present our numerical results on the phase transition and the thermodynamic properties of the model. As it was the case in the one-component scalar model studied in Ref.~\cite{Marko:2012wc}, we shall find that in the chiral limit, the transition, when it occurs, is of the second order type. We shall show this explicitly by monitoring the variation of the order parameter. We shall also determine some critical exponents as well as thermodynamical observables. Before doing so, we discuss the physical parametrization of the model, relevant, when $N=4$, for the discussion of light meson properties.

%%%
\subsection{The parametrization of the model}
The renormalized $O(N)$ model has three parameters $m_\star^2,\,\lambda_\star$ and $h$ ($h=0$ in the chiral limit) and a renormalization scale $T_\star.$ Being the solution of the gap equations at $\phi=0$ and $T=T_\star$, $m_\star^2$ is positive, and since we want the bare couplings to be positive, we need to restrict to $\lambda_\star>0$ (in addition to $\Lambda<\Lambda_{\rm p}$). Not all the $4$-uples $(m_\star,\lambda_\star,h,T_\star)$ correspond to different physical systems. First of all, renormalization group invariance implies that given two values for the renormalization scale $T_\star$, there exists a renormalization group transformation that maps two sets of values for $m_\star$ and $\lambda_\star$ in such a way that the physical predictions are the same. This is rigorously true in the exact theory where no approximation is considered but it needs not be the case in a given truncation of the $\Phi$-derivable potential and the $T_\star$ dependence of the physical results needs to be investigated, see below. 

Another source of redundancy is provided by dimensional analysis, since knowing the values of the physical observables of a system represented by $(m_\star,\lambda_\star,h,T_\star),$ one can very easily deduce the values of the same physical observables for a rescaled system represented by $(\alpha m_\star,\lambda_\star,\alpha^3h,\alpha T_\star),$ where all dimensionful quantities are rescaled by $\alpha$ to the appropriate power. In contrast to renormalization group invariance, this redundancy is present at any level of truncation and it is therefore convenient to get rid of it by working exclusively with dimensionless parameters. For instance, below we will be interested in the value of the order parameter at $T=0,$ which is a function $\bar\phi_0=\bar\phi_0(m^2_\star,\lambda_\star,h;T_\star)$ of mass dimension one (the label $0$ emphasizes that the given quantity is computed at $T=0$). Using simple dimensional analysis, we deduce that
\beq
\bar\phi_0/T_\star=\bar\phi_0(m^2_\star/T^2_\star,\lambda_\star,h/T^3_\star;1)\,.\ \ \
\label{Eq:scaled_phi0}
\eeq
Similar expressions can be obtained for the rescaled curvature masses $\hat M_{\rm L,0}/T_\star$ and $\hat M_{\rm T,0}/T_\star$ that are also needed below. The use of rescaled variables $m_\star^2/T^2_\star$ and $h/T^3_\star$ as parameters, is more suitable for numerical calculations for only dimensionless numbers are used and according to Eq.~\eqref{Eq:scaled_phi0} we can replace $T_\star$ by 1 in the numerical code.\\

In principle, the parameters can be fixed by equating quantities computed at zero temperature with their experimental values. Our choice is to relate $\bar\phi_0$ with the pion decay constant $f_\pi$ and the curvature masses $\hat M_{\rm T,0}$ and $\hat M_{\rm L,0}$ with the mass of the pion and sigma particles, $m_\pi$ and $m_\sigma$ respectively. We decided to use those masses for they reflect the best symmetry of the theory whereas, as discussed in Sec.~\ref{subsec:equations}, the gap masses violate the Ward identities associated to the $O(N)$ symmetry, e.g. Goldstone's theorem. However, the choice of curvature masses for parametrization is questionable, since usually the measured physical masses are the pole masses. In this work, we do not have access to the spectral functions and therefore we assume implicitly that the pole masses are not so far from the curvature modes (this of course would deserve further investigation).

One way to proceed would be to choose a value of $T_\star$ and equate $\bar\phi_0$ in Eq.~(\ref{Eq:scaled_phi0}) to $f_\pi:$
\beq
\bar\phi_0=T_\star\bar\phi_0(m_\star^2/T^2_\star,\lambda_\star,h/T^3_\star;1)\overset{!}{=}f_\pi
\label{Eq:fpi_cond}
\eeq
and similarly for $m_\sigma$ and $m_\pi$. This would define a point in the parameter space $(m^2_\star/T^2_\star,\lambda_\star,h/T^3_\star)$. By changing the value of $T_\star$ without changing the values of $f_\pi$, $m_\sigma$ and $m_\pi,$ we would then follow a line of constant physics. One difficulty with this approach is that our renormalization procedure requires in the chiral limit the temperature $T_\star$ to be necessarily in the symmetric phase and thus for a given set of physical values of $f_\pi$, $m_\sigma$ and $m_\pi$ there is a minimal possible value for $T_\star$ which we do not know {\it a priori}. Another difficulty is that the sigma mass is not known exactly, as according to Ref.~\cite{Beringer:1900zz} $m_\sigma\in(400,550){\rm\,MeV}$ and, based on large-$N$ studies \cite{Patkos:2002xb, Andersen:2004ae}, one may have concerns whether in our approximation $\hat M_{\rm L,0}$ turns out to be large enough.\footnote{A maximal value of the sigma pole mass was observed in these studies. This can be seen in Figure~2 of Ref.~\cite{Patkos:2002xb}, and a formula determining the maximal value was derived in Ref.~\cite{Andersen:2004ae}. The renormalization scale used to fix the coupling constant differs in the two references.} Hence, instead of trying to fix the parameters by picking up some arbitrary value for the sigma mass in the range given above, our procedure is to scan an appropriately large part of the space $(m_\star^2/T_\star,\lambda_\star,h/T_\star^3)$ and determine at each point $\bar\phi_0/T_\star$ using Eq.~(\ref{Eq:scaled_phi0}) and $\hat M_{{\rm T},0}$ and $\hat M_{\rm L,0}$ from similar relations. At each point of the investigated parameter space we require  $\bar\phi_0=93{\rm\,MeV},$ which fixes $T_\star$ according to Eq.~(\ref{Eq:fpi_cond}) and allows us to determine $\hat M_{{\rm T},0}$ and $\hat M_{\rm L,0}.$ We then keep only those points which satisfy $\hat M_{\rm T,0}=138\pm1.38{\rm\,MeV}$ and allow for the decay of the sigma particle into two pions by requiring $\hat M_{\rm L,0}>2\hat M_{\rm T,0}.$ A one percent tolerance is allowed in the value of $\hat M_{\rm T,0}$ in order to guarantee a sufficient number of points, even when the parameter space is not densely sampled. In the chiral limit, there is no constraint on $\hat M_{\rm T,0},$ because this vanishes due to Goldstone's theorem (see the discussion below Eq.~\eqref{Eq:curvature_masses}), and hence the constraint on the sigma mass is lifted as well. Another difference is that the value for the pion decay constant in the chiral limit is $f_\pi^{h=0}=88{\rm\,MeV}$~\cite{Gasser:1983yg} instead of $f_\pi=93{\rm\,MeV}$ used at $h\ne 0.$ 

Since by construction all points that we keep are such that $\bar\phi_0$ and $\hat M_{{\rm T},0}$ are fixed, the iso-$\hat M_{\rm L,0}$ curves are ``lines of constant physics''. We use quotation marks because, as already mentioned, in a given truncation, we expect physical quantities to vary slightly as we move along such a line, that is as we change $T_\star$ for fixed $f_\pi,$ $m_\sigma$ and $m_\pi$.\footnote{Also, even though in the exact theory, the lines of constant physics should have a constant $h$, this does not need to be the case in a given truncation.} Along such a line we can determine in particular $T_{\rm pc}$ ($T_{\rm c}$ at $h=0$) from the inflection point of the $\bar\phi(T)$ curve as described in Sec.~\ref{sec:numerics} and plot its dependence with respect to $T_\star$. We can apply the same strategy for any other physical quantity and study its physical dependence, see Sec.~\ref{ss:thermo} for a discussion concerning the pressure. Note finally that, even though all points in parameter space correspond by construction to a given value of $\bar\phi_0,$ we could access other values of $\bar\phi_0$ (if our model would apply to other physical systems) by using dimensional analysis and changing the corresponding value of $T_\star.$ Of course, all other dimensionful quantities would be scaled by the same quantity.\\

\begin{figure}
%\vspace*{0.5cm}
\includegraphics[width=0.48\textwidth]{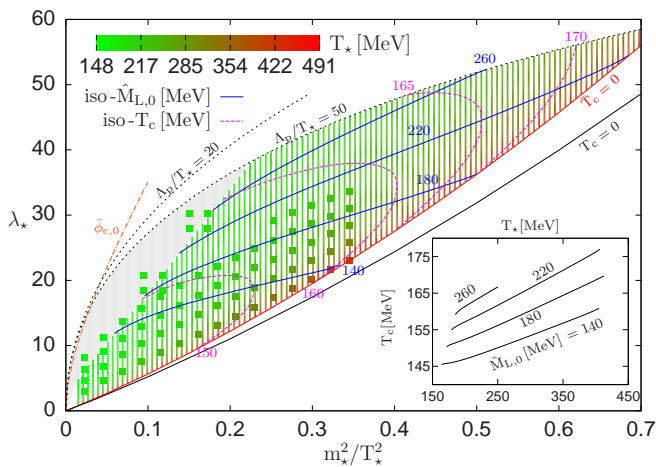}
\caption{Parametrization in the chiral limit ($h=0$). The scanned region is  bounded by the $\Lambda_{\rm p}/T_\star=50$ (upper) and $\bar T_c=0$ (lower) curves. The $\bar\phi_{c,0}$ curve is only present in the hybrid approximation, in which case the grey region is excluded for a reason explained in the text. The points which form vertical lines are obtained in the hybrid approximation, while the squares denote the solution of the full two-loop approximation. The iso-$T_{\rm c}$ and the iso-$\hat M_{\rm L,0}$ curves are obtained in the hybrid case. The palette shows the value of the renormalization scale $T_\star.$ The inset shows the variation of $T_{\rm c}$ with $T_\star$ along iso-$\hat M_{\rm L,0}$ curves. \label{Fig:chiralParams}}
\end{figure}

The result of the parametrization in the chiral limit is shown in Figure~\ref{Fig:chiralParams} in an almond-shaped range of the parameter space encountered already in Refs.~\cite{Reinosa:2011ut,Marko:2012wc}. The $\Lambda_{\rm p}/T_\star=50$ curve can be easily obtained from Eq.~\eqref{Eq:Lp} or Eq.~\eqref{Eq:Lp_approx}, the $\bar T_{\rm c}=0$ curve is given by Eq.~\eqref{Eq:blc_line}, while the $T_{\rm c}=0$ curve is obtained by solving Eq.~\eqref{Eq:lc_line} using \eqref{Eq:M_Labert}. The points investigated in the two-loop case are shown in Figure~\ref{Fig:chiralParams} by squares in order to distinguish them from those used in the hybrid approximation which populate more densely the studied region and appear in the form of vertical lines. 

\begin{figure*}[!t]% environment for wide figures
\includegraphics[width=0.49\textwidth]{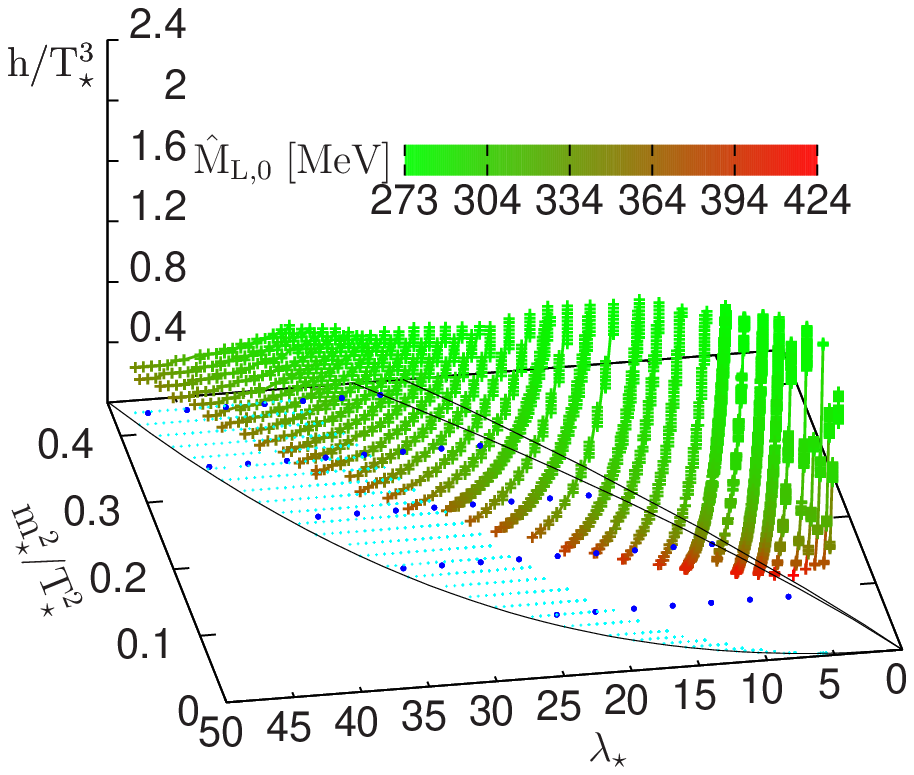}\includegraphics[width=0.49\textwidth]{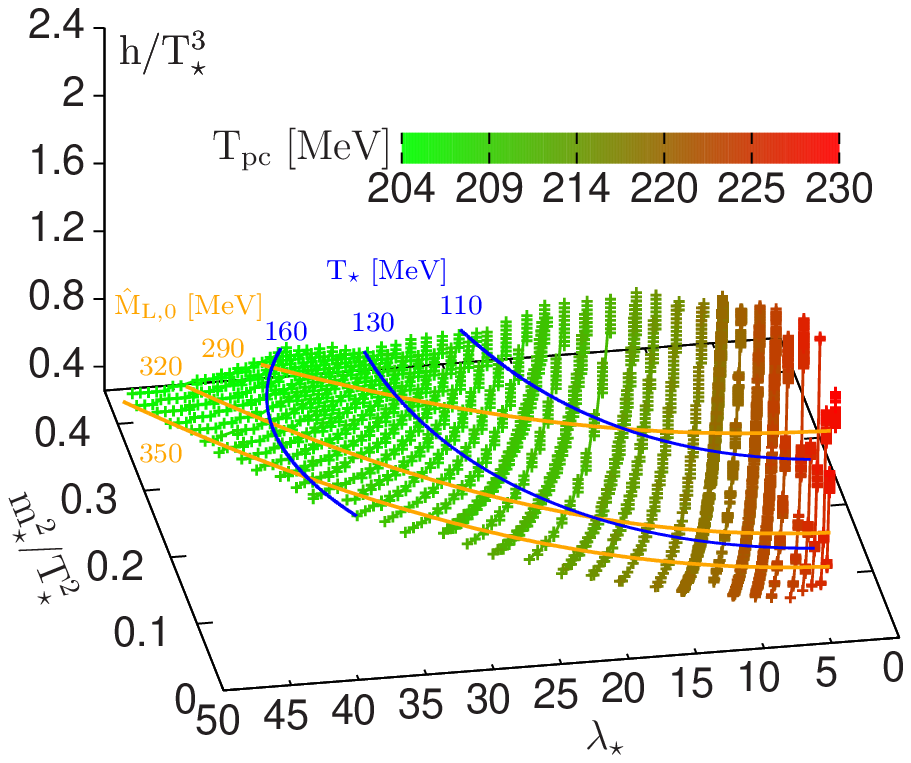}
\caption{Parametrization at $h\ne 0$ in the hybrid case. The location of the investigated points relative to the characteristic curves is indicated with smaller size points in the $(\lambda_\star,m_\star^2/T_\star^2)$ plane of the figure in the left panel. The points having bigger size indicates the parameters used in Figure~\ref{Fig:diff_2l_vs_hybrid} to compare the result of the two-loop and hybrid approximations. The smaller size points satisfy the two criteria $\hat M_{\rm T,0}=138\pm 1.38$~MeV and $\hat M_{\rm L,0}\ge 2 \hat M_{\rm T,0}.$ The value of the renormalization scale $T_\star$ is indicated on the figure in the right panel. The two palettes show the values of $\hat M_{\rm L,0}$ and $T_{\rm pc},$ respectively.\label{Fig:physParams}}
\end{figure*}

In the hybrid approximation, the region to the left of the $\bar\phi_{\rm c,0}$ curve is excluded at $h=0$ because, as discussed in Sec.~\ref{ss:curves}, the model cannot be solved at $T=0$ in that region. Actually, the presence of this line, along which $\bar M_{\rm T,0}=0,$ invalidates the use of the hybrid approximation in the chiral limit in a relatively large region of the parameter space, the grey region of Figure~\ref{Fig:chiralParams}. This is because as one enters this region, by decreasing for example $m_\star^2/T_\star^2$ at fixed $\lambda_\star,$ $\hat M_{\rm L,0}$ increases very abruptly. Such a huge sensitivity to the parameters alone raises suspicion concerning the applicability of the approximation, but in our case one can check explicitly that the results of the hybrid approximation deviate in this case from those obtained in the full two-loop approximation. The right boundary of this region is given by the points where the relative change of $\hat M_{\rm L,0}$ compared to the two-loop approximation equals 3\%. Apart from this excluded region, the results obtained in the two approximations are very close to each other. The value of $\hat M_{\rm L,0}$ (sigma mass) which can be reached is relatively low, less than 300~MeV, and the critical temperature is in the range $[135,190]$~MeV. The scale $T_\star,$ at which the renormalization and consistency conditions are imposed varies in a relatively large interval. Once determined, it allows us to access the value of the Landau pole $\Lambda_{\rm p}$ in physical units and one sees that, in the range of the parameter space where the sigma mass is the largest, $\Lambda_{\rm p}>8.5$~GeV. The inset shows the dependence of $T_{\rm c}$ on $T_\star$ along a line of constant physics. Interestingly, as one goes to larger values of $m_\star^2/T_\star^2$ along these lines, that is as one increases $T_\star$, the dependence becomes linear. \\ 
%Due to the limitation imposed by the Landau scale this can be seen only for relatively low sigma masses.

\begin{figure}[!t]
\includegraphics[width=0.49\textwidth]{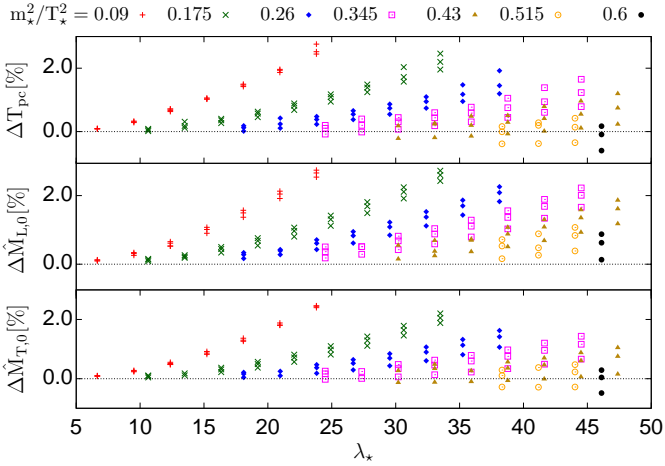}
\caption{Comparison between the two-loop and hybrid approximations based on the relative change of the $T=0$ curvature masses used for parametrization and the pseudocritical temperature ($\Delta Q[\%]=100(Q_{\rm two-loop}/Q_{\rm hybrid}-1)$). Different point types denote different values of $m^2_\star/T_\star^2$ and for each corresponding value of $\lambda_\star$ there are points at three values of $h/T_\star^3:$ 0.5, 1.5, 2.5, which generally increase from bottom to top. \label{Fig:diff_2l_vs_hybrid}}
\end{figure}

The result of the parametrization when $h\ne0$ is shown in Figure~\ref{Fig:physParams} for the hybrid approximation. Compared to the chiral limit we see an increase in the value of $\hat M_{\rm L,0}$ and of the pseudocritical transition temperature $T_{\rm pc}$ and a significant decrease in the value of the renormalization scale $T_\star.$ For fixed $m^2_\star/T_\star,$ larger values of $\hat M_{\rm L,0}$ can be achieved for higher $\lambda_\star,$ that is allowing the Landau pole to come closer to the physical scales. We note that a similar figure could be obtained in the two-loop approximation, but with a significantly increased numerical effort. In the hybrid case the code is much faster than in the two-loop case and hence one can run it for a much larger number of points of the parameter space. We have tested on a good number of points of the scanned region, even those not satisfying $\hat M_{\rm L}>2\hat M_{\rm T},$ that for a given set of the parameters the two-loop results for $\hat M_{\rm L,0},$ $\hat M_{\rm T,0}$ and $T_{\rm pc}$ are within 3\% of the values obtained in the hybrid approximation. This is shown in Figure~\ref{Fig:diff_2l_vs_hybrid}, where the general tendency is that at fixed $m^2_\star/T_\star^2$ both $\lambda_\star$ and $h/T_\star^3$ tend to increase the difference, so that the largest difference is obtained at the largest $\lambda_\star$ and $h/T_\star^3,$ and that this largest difference decreases with increasing $m^2_\star/T_\star^2.$\\ 

\begin{figure}[!t]
\includegraphics[width=0.49\textwidth]{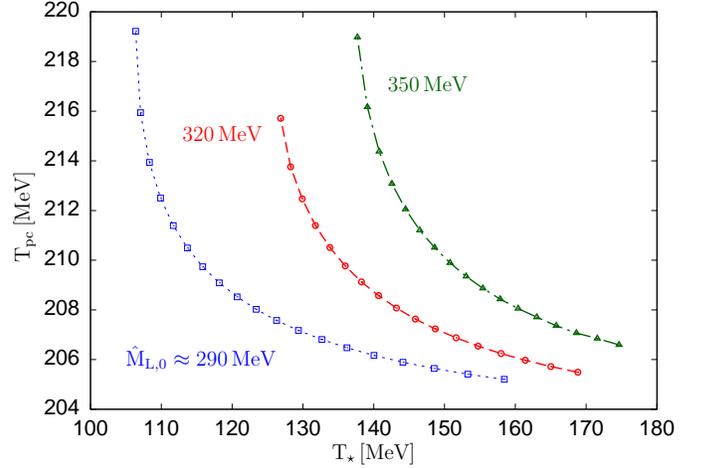}
\caption{The dependence of the pseudocritical temperature $T_{\rm pc}$ on the renormalization scale $T_\star$ in the hybrid approximation at $h\ne0$ along different lines of constant physics specified by the value of $\hat M_{\rm L,0}.$\label{Fig:Tpc_v_Ts}}
\end{figure}

Figure~\ref{Fig:Tpc_v_Ts} shows the variation of the pseudocritical temperature with the renormalization scale $T_\star$ determined during parametrization in the physical case and in the hybrid approximation. The lines of the figure belong to different curves in the $(m_\star^2/T_\star^2,\lambda_\star,h/T_\star^3)$ parameter space selected by different values of $\hat M_{\rm L,0},$ each of them being a line of constant physics. One sees that the $T_\star$ dependence is less than 10\%. %The nature of this dependence is changed compared to the one that can be seen in the chiral limit in the inset of Figure~\ref{Fig:chiralParams}. However, i
In units of $T_\star,$ both $T_{\rm c}$ and $T_{\rm pc}$ decrease for increasing $T_\star$ and for large values of $T_\star$ one can fit (up to possible logs) $a+b/x$ on $T_{\rm c}/T_\star$ and $T_{\rm pc}/T_\star.$ In both cases $b>0,$ but in the chiral limit $a>0,$ while for $h\ne0$ one has $a<0,$ which accounts for the increase of $T_{\rm c}$ and decrease of $T_{\rm pc}$ seen in Figure~\ref{Fig:chiralParams} and in Figure~\ref{Fig:Tpc_v_Ts} for a given line of constant physics and for large $T_\star.$ We expect $|a|$ to diminish as we increase the order of truncation.\\

%%%
\subsection{On the sigma mass\label{ss:sigma}}

\begin{figure*}% environment for wide figures
\includegraphics[height=0.35\textheight,width=0.85\textwidth]{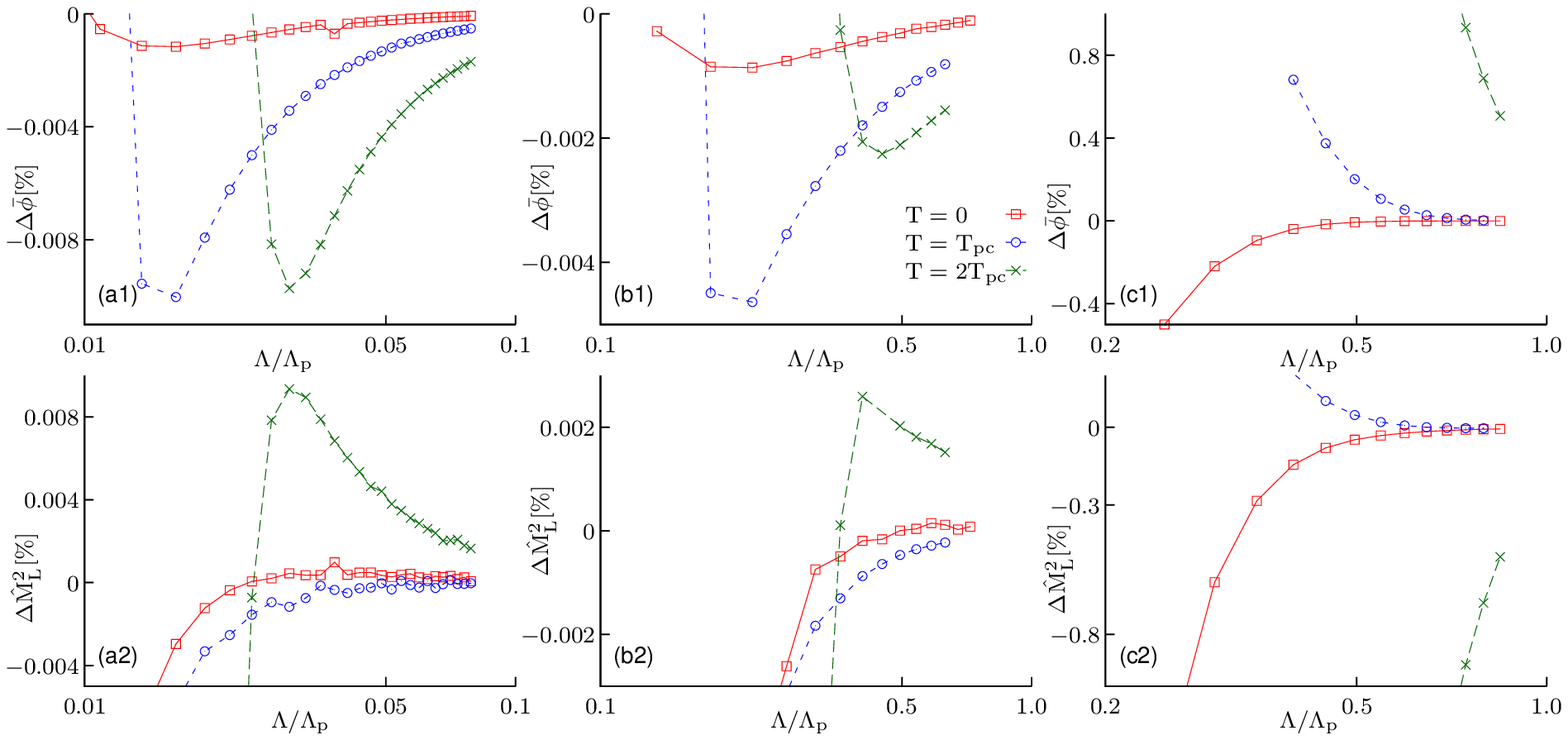}
\caption{The cutoff dependence of the relative change of $\bar\phi$ (upper raw) and that of $\hat M_{\rm L}^2$ (lower raw) obtained in the two-loop approximation at different temperatures: $T=0,$ $T=T_{\rm c}$ %(for each parameter set it's own value, however the same for all cutoffs at a given parameter set), 
and $T=2T_{\rm c}.$ The different parameter sets are: (a) $m_\star^2/T_\star^2=0.124,$ $\lambda_\star=22.28,$ $h/T_\star^3=1.775$ for which $\hat M_{\rm L,0}\approx280$~MeV, $\Lambda_{\rm p}\approx186$~GeV and $T_\star\approx101$~MeV; (b) $m_\star^2/T_\star^2=0.04,$ $\lambda_\star=17.39,$ $h/T_\star^3=0.6,$ for which $\hat M_{\rm L,0}\approx360$~MeV, $\Lambda_{\rm p}\approx16.2$~GeV and $T_\star\approx146$~MeV; (c) $m_\star^2/T_\star^2=0.04,$ $\lambda_\star=19.2,$ $h/T_\star^3=0.38$ for which $\hat M_{\rm L,0}\approx465$~MeV, $\Lambda_{\rm p}\approx3.35$~GeV and $T_\star\approx167$~MeV. The given $\hat M_{\rm L,0}$ and $T_\star$ values correspond to the largest $\Lambda$ point of each set. The discretization is characterized by $N_\tau=512$ and $N_s=3\times2^{10}$ except for the points of set (a) at $T=0$ for $\Lambda/\Lambda_{\rm p}>0.04$ where $N_\tau=3\times512$ was used. The step $\Delta L/T_\star$ was 5 for the cases (a) and (b) and 1 for case (c). \label{Fig:CO_dep}}
\end{figure*}

The parametrization reveals that there is a large region of the parameter space where a separation of scale occurs in the sense that the physical scales are much lower than the cutoff, which in turn is much smaller than the scale of the Landau pole $\Lambda_{\rm p}.$ In this case the solution of the model is practically insensitive to the cutoff used, as it was the case for $N=1$ in Ref.~\cite{Marko:2012wc}, where the cutoff dependence was thoroughly investigated. We have also seen that the value of the zero temperature sigma mass defined through $\hat M_{\rm L,0}$ increases with increasing $\lambda_\star$. We have reached values of sigma masses which are larger than the maximal value of the sigma pole mass found within the large-$N$ approximation in Ref.~\cite{Patkos:2002xb}, which in the chiral limit is $m_\sigma\approx 328$~MeV obtained for a coupling $\lambda\approx311$ and a renormalization scale of $M_0\approx334$~MeV and $m_\sigma\approx 362$~MeV in the $h\ne0$ case, obtained for $\lambda\approx386$ and $M_0\approx381$~MeV. The scale of the Landau pole in these cases is approximately $1853$~MeV and $1150$~MeV, respectively. In Ref.~\cite{Andersen:2004ae}, where the renormalization scale and the value of the coupling were chosen differently, a higher value of the sigma pole mass of around 433~MeV was reported. However, in that case, the scale of the Landau pole was only 720~MeV which prevented calculations above $T\approx 50$~MeV. 

We investigate now what happens in our case with the scale of the Landau pole, which in view of Eq.~\eqref{Eq:Lp} decreases with $\lambda_\star$ when all the other parameters are kept fixed, if a more realistic parametrization of the model is required, in which $m_\sigma\in[440,470]$~MeV to conform to recent dispersive analyses of more precise $\pi\pi$ scattering data (see Ref.~\cite{Beringer:1900zz} and for a recent review Ref.~\cite{Pelaez:2013jp}, in particular its Figure~3). To this end, we have chosen different values of $m_\star^2/T_\star^2$ and increased the value of $\lambda_\star$ in the range between the $\Lambda_{\rm p}/T_\star=50$ and $\Lambda_{\rm p}/T_\star=20$ curves of the $(m_\star^2/T_\star^2,\lambda_\star)$-plane, shown in Figure~\ref{Fig:chiralParams}. It turns out that in the two-loop approximation it is possible to reach with the parametrization procedure described in the previous subsection values of the $\hat M_{\rm L,0}$ in the desired range. For instance, we obtain $\hat M_{\rm L,0}\approx465$~MeV and $T_\star\approx167$~MeV for $m_\star^2/T_\star^2=0.04,$ $h/T_\star^3=0.38,$ $\lambda_\star=19.2,$ and $\hat M_{\rm L,0}\approx445$~MeV and $T_\star\approx171$~MeV for $m_\star^2/T_\star^2=0.124,$ $h/T_\star^3=0.355,$ $\lambda_\star=32.476.$ In these cases the scale of the Landau pole remained at least seven times larger than the largest mass scale given by $\hat M_{\rm L}$, that is $\Lambda_{\rm p}\simeq 3.4$~GeV. 

The interesting question is whether the scale of the Landau pole is high enough for the result not to depend too much on the value of the cutoff $\Lambda.$ In order to study the sensitivity of the results on $\Lambda$ we monitored the cutoff dependence of the relative change $\Delta Q(\Lambda)=(Q(\Lambda+\Delta\Lambda)/Q(\Lambda)-1)$ of a given physical quantity $Q$. If this quantity is regarded as a sequence for discrete values of $\Lambda,$ then in the ideal case where the convergence occurs the relative change not only tells us how sensitive is the physical quantity on the cutoff at some value $\Lambda,$ but also how close it is to the convergent value. This is because if at some cutoff $\Lambda$ the value $|\Delta Q(\Lambda)|<10^{-n},$ then one can say that $Q(\Lambda)$ is within $10^{-n+2}$~\% from the converged value of the physical quantity $Q.$ The problem is, of course, that strictly speaking the convergence would occur as $\Lambda\to \infty,$ but generally one cannot go above the Landau pole. Therefore, what is of practical relevance is whether the $Q$ shows a plateau as a function of $\Lambda$ below the scale of the Landau pole. We investigate this in Figure~\ref{Fig:CO_dep} for several parameter sets using the quantities $\bar\phi$ and $\hat M^2_{\rm L}$ at different temperatures (the relative change is shown in percentage). One can see that we are closest to a plateau if the scale of the Landau pole is high and the temperature is low. The variation of the relative change with the cutoff shows that even when the scale of the Landau pole is approximately seven times larger than $\hat M_{\rm L,0},$ for practical purposes the result can be considered compatible with a cutoff independent result, at least for temperatures not too large with respect to $T_{\rm pc}.$ This result should however be interpreted with a pinch of salt since the fact that the plateau observed in Figure~\ref{Fig:CO_dep} extends up to the Landau scale is related to the fact that the physical quantities do not diverge at this scale, only the bare couplings. In higher order approximations where, due to a negatively quadratic growth of the self-energy at large frequency/momentum or to vertex-type resummations, one expects physical quantities to diverge at $\Lambda_{\rm p},$ it is less probable that a plateau can appear if the Landau scale is too low.\footnote{Some explicit calculation using an educated example reveals that, if $\Lambda_{\rm p}$ is large enough, a plateau develops and even extends up to the very vicinity of $\Lambda_{\rm p}$. However, as $\Lambda_{\rm p}$ is decreased, the plateau fades away.}

%%%
\subsection{Phase transition\label{ss:PD}}

\begin{figure}
\begin{center}
\includegraphics[width=0.49\textwidth,angle=0]{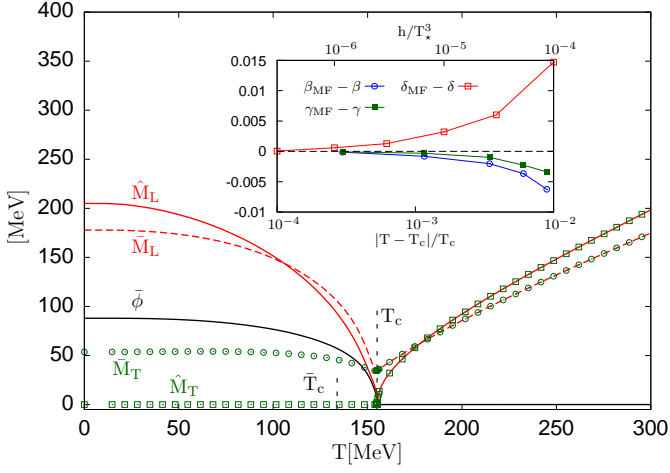}
\caption{Illustration of the second order nature of the phase transition in the chiral limit within the two-loop approximation. The inset shows the convergence of the static critical exponents to the mean-field values, as the upper boundary of the fitting range is decreased. The parameters are: $m_\star^2/T_\star^2=0.124,$ $\lambda_\star=22.277,$ while the discretization is characterized by $\Lambda/T_\star=100,$ $N_s=3\times 2^{10}$ and $N_\tau=512$ for $T>77$~MeV and $N_\tau=2\times 2^{10}$ for $T\le77$~MeV. We also defined $\bar M_{\rm L/T}\equiv \bar M_{\rm L/T}(0,\Delta k).$\label{Fig:criticality}}
\end{center}
\end{figure}

In the chiral limit, in both the two-loop and the hybrid approximations, the model undergoes a second order phase transition for those parameters of the $(m^2_\star/T_\star^2,\lambda_\star)$ plane, which are located above the $T_{\rm c}=0$ line of Figure~\ref{Fig:chiralParams}. This is illustrated within the two-loop approximation in Figure~\ref{Fig:criticality}, where we show the temperature evolution of the field expectation value, curvature masses and gap masses at the lowest available momentum. The inset shows that the three numerically determined critical exponents are compatible with the values $\beta=1/2,$ $\gamma=1$ and $\delta=3$: thus the critical behavior of the system is characterized by mean-field-type critical exponents at this level of approximation. This is expected, since they were already found to be of the mean-field type in the one-component case in Ref.~\cite{Marko:2012wc}. 
%The critical exponent $\beta$ was obtained by fitting $(T_{\rm c}-T)^\beta$ to the values of $\bar\phi$ determined in the broken phase, while $\gamma$ was obtained by fitting the inverse curvature of the potential determined in the symmetric phase, $(\hat M^2_{\bar\phi=0})^{-1},$ with $(T-T_{\rm c})^{-\gamma}.$ The critical exponent $\delta$ was obtained at the critical value of the temperature, by fitting $h^{1/\delta}$ to the solution $\bar\phi$ of the field equation determined for decreasing values of the external field $h.$  

One also sees in Figure~\ref{Fig:criticality} that $\hat M_{\rm T}$ fulfills the requirement of Goldstone's theorem discussed in Sec.~\ref{subsec:equations} around Eq.~\eqref{Eq:curvature_masses}, as it vanishes in the broken phase and becomes degenerate with  $\hat M_{\rm L}$ in the symmetric phase. However, as a result of the truncation of the 2PI effective action, Goldstone's theorem is violated by $\smash{\bar M_{\rm T}(K=0)}$ (approximated numerically by $\bar M_{\rm T}(0,\Delta k)$) which is rather large since at small temperatures it is larger than $\bar\phi/2.$ We note however that $\bar M_{\rm T}(0,\Delta k)$ is the smallest scale among $\bar M_{\rm T}(0,\Delta k)$, $\bar M_{\rm L}(0,\Delta k)$, $\hat M_{\rm L}$ and $\bar\phi$ and that the size (in MeV) of the violation of Goldstone's theorem is quite constant with the temperature. These observations give good hope that higher order corrections will reduce uniformly the violation of Goldstone's theorem by the transverse gap mass. The restoration of Goldstone's theorem is expected because $\bar M=\hat M$ in the absence of approximations. Our results indicate that the restoration could happen uniformly with the temperature. At large temperature both the degenerate curvature and gap masses increase, but a gap remains between them. This reflects the fact that the two-loop approximation is such that $\delta^2\Gamma_{\rm int}/\delta\phi_a\delta\phi_b|_{\phi=0}\ne 2\delta \Gamma_{\rm int}/\delta G_{ab}|_{\phi=0},$ where $\Gamma_{\rm int}[\phi,G]$ contains all the 2PI graphs contracted with the vertices of the shifted action $S[\varphi\to\varphi+\phi]$ (see Ref.~\cite{Berges:2005hc} for details).\\

\begin{figure}
\begin{center}
\includegraphics[width=0.49\textwidth,angle=0]{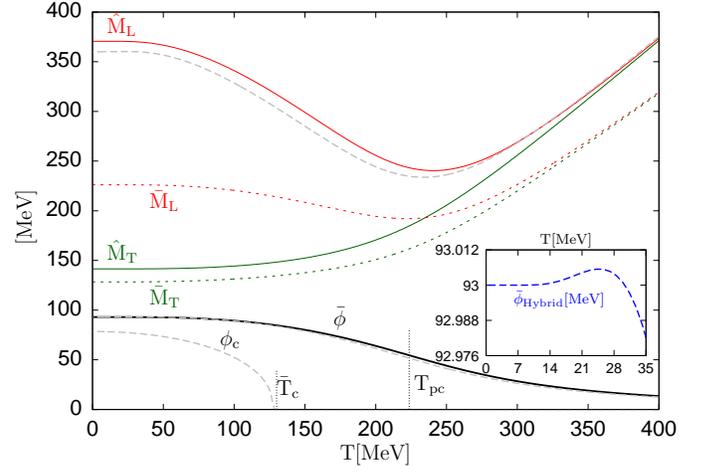}
\caption{The temperature evolution of the order parameter, the curvature and gap masses in the two-loop approximation where $\bar M_{\rm L/T}\equiv \bar M_{\rm L/T}(0,\Delta k).$ For $\bar\phi$ and $\hat M_{\rm L}$ we also show for comparison the curves obtained in the hybrid approximation, in which case we show the critical value of the field below which, at a given temperature, the effective potential is not accessible. The parameters are: $m_\star^2/T_\star^2=0.04,$ $\lambda_\star=17.39,$ $h/T_\star^3=0.6.$ The discretization used in the two-loop case is characterized by $\Lambda/T_\star=55,$ $N_s=3\times 2^{10}$ and $N_\tau$ was increased for decreasing temperature from 512 used at $T\ge40$~MeV to $2\times 2^{10}$ for $T\in [25,40]$~MeV and $4\times 2^{10}$ for $T\le 25$~MeV.\label{Fig:masses}}
\end{center}
\end{figure}

In the physical case ($h\ne 0$) the thermal transition is of an analytic crossover type. The temperature evolution of the order parameter is presented in Figure~\ref{Fig:masses} for a set of parameters at which $\hat M_{\rm L,0}\approx 360$~MeV. The solid line is obtained in the two-loop approximation, while the barely distinguishable dashed line is obtained in the hybrid approximation at the same values of the parameters. The inset shows that in the hybrid approximation $\bar\phi(T)$ is not a monotonous function of the temperature, for it shows a maximum at some value of the temperature. This reflects an inconsistency of the hybrid approximation because, as one sees in Figure~\ref{Fig:thermo}, in the temperature range where $\bar\phi(T)>\bar\phi_0,$ the pressure is negative. In Figure~\ref{Fig:masses}, the difference between the two approximations is more visible on the longitudinal curvature mass (the transverse ones differ very little because $\hat M_{\rm T}^2=h/\bar\phi$). The restoration of symmetry at high temperature is reflected by both the curvature and lowest momentum gap masses, as the corresponding longitudinal and transverse components approach each other. As in the chiral case, at large temperature there remains a gap between the curvature and gap masses. We note also the important difference between the values of $\bar M_{\rm L}(0,\Delta k)$ and $\hat M_{\rm L}>\bar M_{\rm L}(0,\Delta k)$ at $T=0$. It is clearly more convenient to use the curvature masses to parametrize the model since they allow to reach higher sigma masses for the same values of the parameters.

%%%
\subsection{Thermodynamics\label{ss:thermo}}

\begin{figure}[!t]
\begin{center}
\includegraphics[width=0.475\textwidth,angle=0]{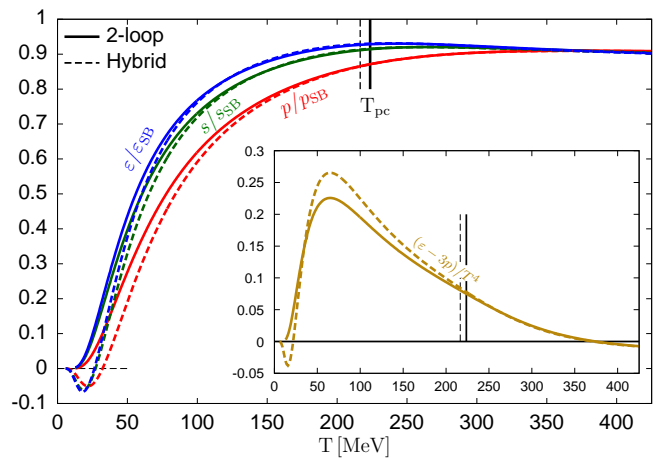}
\caption{The temperature dependence of the pressure, energy density and entropy density normalized to the corresponding Stefan-Boltzmann limit obtained for the same parameter set of Figure~\ref{Fig:masses} in the the two-loop (continuous lines) and hybrid approximations (dashed lines). The inset shows the variation of the interaction measure with the temperature.\label{Fig:thermo}}
\end{center}
\end{figure}

\begin{figure}[!t]
\begin{center}
\includegraphics[width=0.475\textwidth,angle=0]{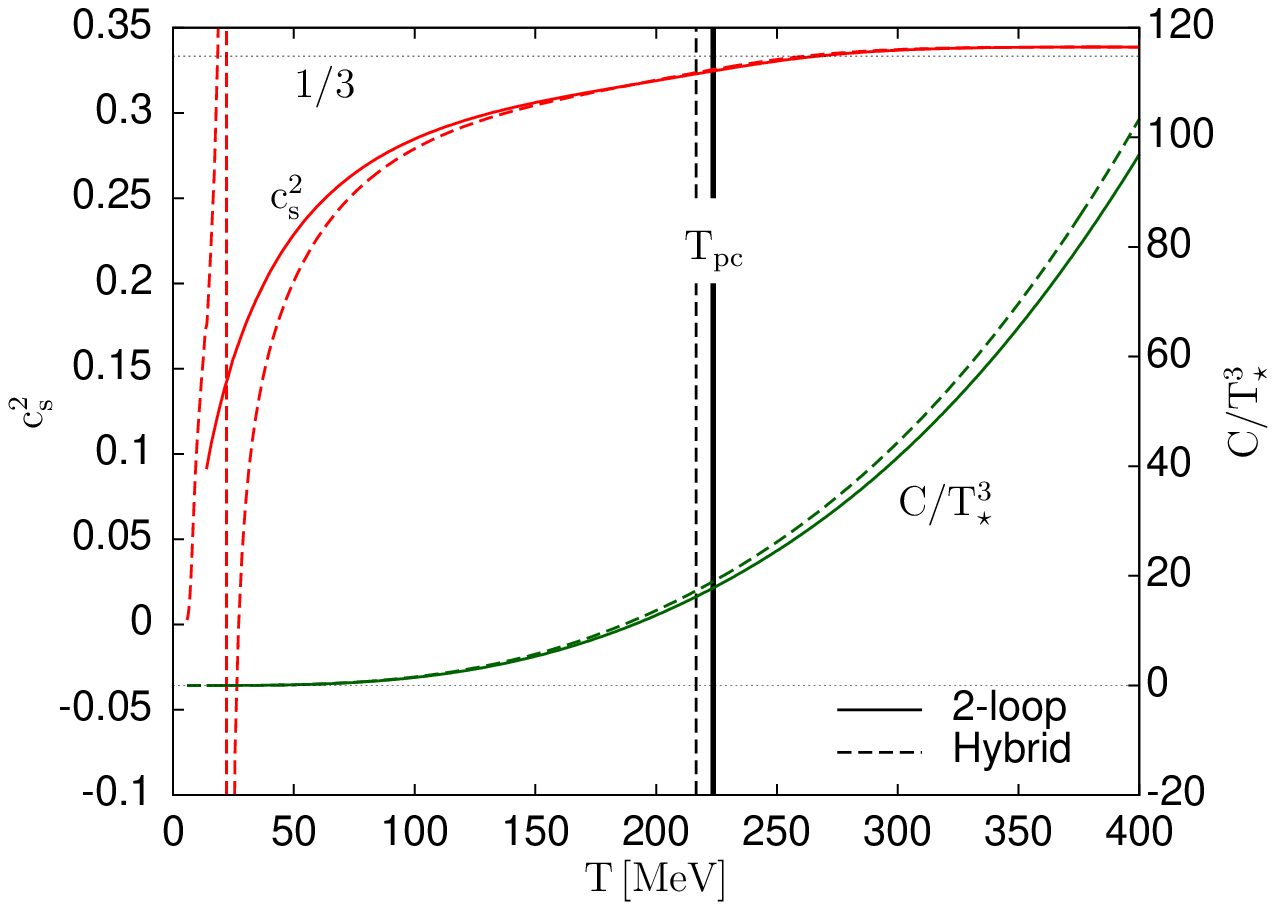}
\caption{The square of the speed of sound $c_{\rm s}^2$ and the heat capacity $C$ obtained for the same parameter set of Figure~\ref{Fig:masses} in the the two-loop (continuous lines) and hybrid approximations (dashed lines). The left and right axes of the plot correspond to $c_{\rm s}^2$ and $C,$ respectively.
\label{Fig:cs2_C}}
\end{center}
\end{figure}

We turn now to the study of the thermodynamic properties of the model. To this end we compute the pressure by subtracting the value of effective potential at the minimum obtained at a given temperature from the value determined at zero temperature as described below Eq.~(81) of Ref.~\cite{Marko:2012wc}. The entropy density $s=dp/dT$ is determined from the pressure through a numerical derivative, while the energy is calculated as $\epsilon=-p+T s.$ Usually these quantities are divided with appropriate powers of the temperature, but we choose to normalize them to the corresponding quantity calculated for an ideal gas of massless particles. As it was the case for $N=1$ these three quantities when rescaled with the corresponding Stefan-Boltzmann limit agree with each other at that temperature where the interaction measure $\Delta = T d(p/T^4)/d T=(\epsilon-3p)/T$ vanishes. As discussed in Ref.~\cite{Marko:2012wc}, this feature follows directly from the equations, and can be seen in Figure~\ref{Fig:thermo}, where we compare the dependence of these quantities on the temperature in the two-loop and hybrid approximations. The curves obtained in the two cases are indistinguishable above the pseudocritical temperature. At small temperature, however, there are visible differences, and more importantly one can clearly see that the hybrid approximation is not consistent from a thermodynamic point of view, since at small temperatures it leads to negative pressure, entropy and energy densities. As already mentioned the temperature region where the pressure is negative is correlated to that where $\bar\phi(T)>\bar\phi_0.$ The inconsistency of the hybrid approximation is displayed also by the heat capacity $C=T d^2p/d T^2,$ which becomes negative for small temperature and by the square of the speed of sound $c_{\rm s}^2=dp/d\epsilon=s/C$ which has a singularity at the temperature for which $C$ vanishes (see Figure~\ref{Fig:cs2_C}). Note, that for the two-loop case the temperature variation of $c_{\rm}^2$ is much milder than the one shown in Figure~8 of Ref.~\cite{Li:2009by} which was obtained in a Hartree approximation which included only the thermal effects and neglected the vacuum ones. This is probably related to the fact that in our case $\bar M_{\rm L}(T=0)\approx 225$~MeV is smaller than the smallest value of the corresponding mass used there.

It is visible in  Figure~\ref{Fig:thermo} that at high temperature the pressure normalized to the Stefan-Boltzmann limit decreases with the temperature. This is the consequence of the fact that, as one can see in Figure~\ref{Fig:masses}, at high $T$ the masses of the excitation grow linearly with $T$ and therefore a high temperature expansion is less and less accurate with increasing $T.$ 

In the two-loop approximation we also tested the dependence of the pressure on the renormalization scale $T_\star,$ by choosing two points in the parameter space which belong to a line of constant physics, along which $\hat M_{\rm L/T,0}(0,\Delta k)$ and $\bar\phi_0$ are constant, and for which the difference in $\bar M_{\rm T,0}(0,\Delta k)$ was maximal, that is around 10\%. The difference in the value of $T_\star$ corresponding to these two points was around 10\% and although the difference of $\bar M_{\rm L,0}(0,\Delta k)$ was around 30\%, the maximal difference in the pressure was around 10\% and was observed at temperatures smaller than $T_{\rm pc}.$

%%%%%
\section{Conclusions\label{sec:conclusions}}
We studied numerically the thermal phase transition of the renormalized $O(N)$ model, both in a genuine $\Phi$-derivable approximation in which the effective action is truncated at two-loop level and in a hybrid approximation in which the effective potential and the field equation derived from it are evaluated with a lower level, Hartree-Fock-type transverse (pion) and longitudinal (sigma) propagators. In the first case the self-consistent propagator equations were solved iteratively in Euclidean space using 3D cutoff regularization and the method of Ref.~\cite{Marko:2012wc}, which by a combination of adaptive numerical integration and fast Fourier transforms ensures a very accurate evaluation of the convolution-type integrals. In the hybrid approximation one obtains explicitly finite equations which are much simpler to solve. 

In the chiral limit the phase transition turns out to be of second order in both approximations studied. On the one hand, this means that the higher level truncation considered in this work represents an improvement over the Hartree-Fock approximation which is known to yield a first order phase transition in the chiral limit. On the other hand, we have a clear indication that the important improvement over the Hartree-Fock level occurs in the field equation and is related to the inclusion of the setting-sun diagram. In the case of an explicit breaking of the chiral symmetry the transition is an analytic crossover. 

As long as one is interested in the temperature evolution of the expectation value of the field, curvature and gap masses the hybrid approximation can be regarded as a good approximation of the two-loop $\Phi$-derivable approximation. In the chiral limit this is not true for the entire parameter space, as one has to restrict its application to those parameters where the longitudinal curvature mass does not change abruptly with the parameters. However, the thermodynamic study revealed its inconsistency at small temperatures for it leads to negative pressure, entropy density and energy density. In fact, this feature is also related to the nonmonotonic behavior of the field expectation value at small temperature, where it first increases with increasing temperature.

We have seen that for $N=4$ it is possible to achieve a realistic parametrization of the model, in which the zero temperature sigma mass, obtained as the longitudinal eigenmode of the curvature tensor, could be fixed to values around 460~MeV, while keeping the scale of the Landau pole at around 3.4~GeV. This scale is large enough for the results to be considered practically independent of the cutoff used, at least for the approximation considered here and for temperatures not too large with respect to the crossover temperature. The values of the sigma mass which can be obtained within the two-loop 2PI approximation are larger than those found in the next-to-leading order of the $1/N$ expansion in the 1PI formalism \cite{Patkos:2002xb,Andersen:2004ae}, and the scale of the Landau pole proved also larger. However, in the approximations studied here, there is a significant difference between the curvature masses and the gap masses. It is expected that in approximations where the effective action is truncated at higher orders this discrepancy will diminish and then the question is raised whether this will affect the maximum value of the sigma mass achievable from the curvature mass. In this respect, it will be interesting to investigate whether the possibility of a realistic parametrization of the $O(4)$ model persists in the 2PI formalism at higher order truncation levels and also what will be the case in the linear sigma model with three flavors at two-loop and higher truncations levels. Also, in higher order approximation, especially in those involving vertex-type resummation, as the 2PI-$1/N$ expansion, the presence of the Landau pole is a more severe problem, because it influences the renormalized quantities due to the divergence of the vertex function. In this case the cutoff insensitivity needs a careful reexamination and it is more probable that the scale of the Landau pole has to be kept further away from the physical scales than in the two-loop approximation discussed here.

\acknowledgments{This work was supported by the French-Hungarian collaboration program PHC Balaton No. 27850RB (T{\'E}T\_11-2-2012). G.\ M. and Zs.\ Sz. were supported by the Hungarian Scientific Research Fund (OTKA) under Contract No. K77534 and No. K104292 and they thank A.~Jakov\'ac and A.~Patk{\'o}s for useful discussions.}

\appendix

%%%%%
\section{Four-point functions}\label{app:four-point}
As it is discussed at length in Ref.~\cite{Berges:2005hc} there exist three different definitions of the four-point function which do not match exactly in a given truncation. Here we shall consider these distinct definitions at $\phi=0$ in which case we can use $\bar G^{\phi=0}_{ab}=\delta_{ab}\bar G_{\phi=0}$. One possible definition is\footnote{In the approximation at hand, the kernel, and in turn the four-point function, are momentum independent, hence the absence of momenta in our notation for the kernel and the four-point function.}
\beq
\bar V^{\phi=0}_{ab,cd} & = & \bar \Lambda^{\phi=0}_{ab,cd}-\frac{1}{2}\int_Q^T \bar\Lambda^{\phi=0}_{ab,uv} \bar G^2_{\phi=0}(Q) \bar V^{\phi=0}_{uv,cd}\,,
\eeq
with
\beq
\bar\Lambda^{\phi=0}_{ab,cd} & \equiv & \left.\frac{4\delta^2\Phi}{\delta G_{ab}(K)\delta G_{cd}(Q)}\right|_{\phi=0}\,.
\eeq
The kernel $\bar\Lambda^{\phi=0}_{ab,cd}$ has the structure $\bar\Lambda^{\phi=0}_{ab,cd}=\bar\Lambda^{(A)}_{\phi=0}\delta_{ab}\delta_{cd}+\bar\Lambda^{(B)}_{\phi=0}(\delta_{ac}\delta_{bd}+\delta_{ad}\delta_{bc})$ with $\bar\Lambda^{(A,B)}_{\phi=0}=\lambda^{(A,B)}_0/3N$. The four-point function $\bar V^{\phi=0}_{ab,cd}$ admits the same decomposition  and its components obey the coupled set of equations
\beq
\bar V^{(A)}_{\phi=0} & = & \bar\Lambda^{(A)}_{\phi=0}-\frac{N}{2}\int_Q^T \bar\Lambda^{(A)}_{\phi=0}\bar G^2_{\phi=0}(Q)\bar V^{(A)}_{\phi=0}\nonumber\\
& & -\int_Q^T \bar \Lambda^{(A)}_{\phi=0}\bar G^2_{\phi=0}(Q)\bar V^{(B)}_{\phi=0}\nonumber\\
& & -\int_Q^T \bar \Lambda^{(B)}_{\phi=0}\bar G^2_{\phi=0}(Q)\bar V^{(A)}_{\phi=0}\,,\label{eq:un}\\
\bar V^{(B)}_{\phi=0} & = & \bar\Lambda^{(B)}_{\phi=0}-\int_Q^T \bar\Lambda^{(B)}_{\phi=0}\bar G^2_{\phi=0}(Q)\bar V^{(B)}_{\phi=0}\,.\label{eq:deux}
\eeq
By expanding this system perturbatively, it is pretty obvious that $\bar V^{(A)}$ and $\bar V^{(B)}$ are not equal and thus that $\bar V_{ab,cd}$ is not crossing symmetric. In particular, the divergent part is not crossing symmetric, which explains the need for two distinct bare couplings $\lambda_0^{(A)}$ and $\lambda_0^{(B)}$. In order to solve the system of equations (\ref{eq:un})-(\ref{eq:deux}), it is convenient to consider the combinations $\bar\Lambda^{(C)}_{\phi=0}\equiv N\bar\Lambda^{(A)}_{\phi=0}+2\bar\Lambda^{(B)}_{\phi=0}=\lambda^{(NA+2B)}_0/3N$ and  $\bar V^{(C)}_{\phi=0}\equiv N\bar V^{(A)}_{\phi=0}+2\bar V^{(B)}_{\phi=0}$. It is then easily proven that
\beq
\bar V^{(C)}_{\phi=0}=\bar\Lambda^{(C)}_{\phi=0}-\frac{1}{2}\int_Q^T \bar\Lambda^{(C)}_{\phi=0}\bar G^2_{\phi=0}(Q) \bar V^{(C)}_{\phi=0}\,,
\eeq
which shows that the combination $\bar V^{(C)}_{\phi=0}$ diagonalizes the system (\ref{eq:un})-(\ref{eq:deux}). This diagonalization is in one-to-one correspondence with the one we used in the case of the Hartree gap equations in Sec.~\ref{sec:hybrid}. In fact the combinations of bare couplings $\bar\Lambda^{(B)}=\lambda_0^{(B)}/(3N)$ and $\bar\Lambda^{(NA+2B)}=\lambda_0^{(NA+2B)}/(3N)$ are precisely those which appeared in the diagonalized form of the Hartree-Fock gap equations. We obtain
\beq
\frac{1}{\bar V^{(B)}_{\phi=0}} & = & \frac{3N}{\lambda^{(B)}_0}+{\cal B}[\bar G_{\phi=0}](0)\,,\label{eq:barVB}\\
\frac{1}{\bar V^{(C)}_{\phi=0}} & = & \frac{3N}{\lambda^{(NA+2B)}_0}+\frac{1}{2}{\cal B}[\bar G_{\phi=0}](0)\,,\label{eq:barVC}
\eeq
which we use in the main text to derive the expressions for $\lambda_0^{(B)}$ and $\lambda_0^{(NA+2B)}$ from the renormalization and consistency conditions. A second definition of the four-point function which depends on one momentum\footnote{In principle, a four-point function depends on three independent momenta. The four-point functions we consider here are taken for particular values or configurations of their external momenta and can thus depend on fewer variables.} is
\beq
V^{\phi=0}_{ab,cd}(K) & = & \Lambda^{\phi=0}_{ab,cd}(K)-\frac{1}{2}\int_Q^T \bar\Lambda^{\phi=0}_{ab,uv} \bar G^2_{\phi=0}(Q) V^{\phi=0}_{uv,cd}(Q)\nonumber\\
& = & \Lambda^{\phi=0}_{ab,cd}(K)-\frac{1}{2}\int_Q^T \bar V^{\phi=0}_{ab,uv} \bar G^2_{\phi=0}(Q) \Lambda^{\phi=0}_{uv,cd}(Q)\,,\nonumber\\
\eeq
with
\beq
\Lambda^{\phi=0}_{ab,cd}(K) & \equiv & \left.\frac{2\delta^3\Phi}{\delta G_{ab}(K)\delta\phi_c\delta\phi_d}\right|_{\phi=0}\nonumber\\
& = & \Lambda^{(A)}_{\phi=0}\delta_{ab}\delta_{cd}+\Lambda^{(B)}_{\phi=0}\Big(\delta_{ac}\delta_{bd}+\delta_{ad}\delta_{bc}\Big)\nonumber\\
\eeq
and
\beq
\Lambda^{(A)}_{\phi=0} & = & \frac{1}{3N}\left(\lambda_2^{(A)}-\frac{2}{3N}\,\lambda^2_\star\,{\cal B}[\bar G_{\phi=0}](K)\right),\ \ \ \\
\Lambda^{(B)}_{\phi=0} & = & \frac{1}{3N}\left(\lambda_2^{(B)}-\frac{N+6}{6N}\,\lambda^2_\star\,{\cal B}[\bar G_{\phi=0}](K)\right).\ \ \ %\nonumber\\
\eeq
Once again the appropriate combination of components $NV^{(A)}_{\phi=0}+2V^{(B)}_{\phi=0}$ leads to a system of decoupled equations for $V^{(B)}_{\phi=0}$ and $V^{(C)}_{\phi=0}$ which is suited in particular to extract the expressions for the bare couplings $\lambda_2^{(A,B)}$. According to Ref.~\cite{Berges:2005hc}, the third possible definition of the four-point function is given by
\beq
\hat V^{\phi=0}_{abcd} & = & \left.\frac{\delta^4\gamma}{\delta\phi_a\delta\phi_b\delta\phi_c\delta\phi_d}\right|_{\phi=0}\nonumber\\
& = & \frac{\lambda_4}{3N}\Big(\delta_{ab}\delta_{cd}+\delta_{ac}\delta_{bd}+\delta_{ad}\delta_{bc}\Big)\nonumber\\
& & -\,\frac{1}{2}\int_Q^T \Lambda^{\phi=0}_{ab,uv}(Q)\bar G^2_{\phi=0}(Q)V^{\phi=0}_{uv,cd}(Q)\nonumber\\
& & -\,\frac{1}{2}\int_Q^T \Lambda^{\phi=0}_{ac,uv}(Q)\bar G^2_{\phi=0}(Q)V^{\phi=0}_{uv,bd}(Q)\nonumber\\
& & -\,\frac{1}{2}\int_Q^T \Lambda^{\phi=0}_{ad,uv}(Q)\bar G^2_{\phi=0}(Q)V^{\phi=0}_{uv,bc}(Q)\,.\nonumber\\
\eeq
Using the tensor decomposition of $V_{\phi=0}$, we check that $\hat V^{\phi=0}_{abcd}=\hat V_{\phi=0}(\delta_{ab}\delta_{cd}+\delta_{ac}\delta_{bd}+\delta_{ad}\delta_{bc})$ (in other words $\hat V^{(A)}_{\phi=0}=\hat V^{(B)}_{\phi=0}$ and thus $\hat V_{\phi=0}$ has the crossing symmetry) with
\beq\label{eq:hV}
\hat V_{\phi=0} & = & \frac{\lambda_4}{3N}-\frac{1}{2N}\int_Q^T \Lambda^{(C)}_{\phi=0}(Q)\bar G^2_{\phi=0}(Q) V^{(C)}_{\phi=0}(Q)\nonumber\\
& & -\,2\left(1-\frac{1}{N}\right)\int_Q^T \Lambda^{(B)}_{\phi=0}(Q)\bar G^2_{\phi=0}(Q) V^{(B)}_{\phi=0}(Q)\,,\nonumber\\
\eeq
where we note that the contributions from the $B$ and $C$ components of $\Lambda_{\phi=0}$ and $V_{\phi=0}$ contribute independently to $\hat V_{\phi=0}$. This expression is used in the main text in order to obtain $\lambda_4$.

%%%%%
\section{Hybrid extras}\label{app:hybrid}
In this section we give some expressions encountered in the hybrid approximation and discuss their particular aspects in some details.

\subsection{Expression of $\hat V_{\phi=0}$}
As mentioned in the main text, the four-point function $\hat V_{\phi=0}$ is modified in the hybrid case. Following the same strategy as in the previous section, we arrive at
\beq
\hat V^{\phi=0}_{abcd} & = & \frac{\lambda_4}{3N}\Big(\delta_{ab}\delta_{cd}+\delta_{ac}\delta_{bd}+\delta_{ad}\delta_{bc}\Big)\nonumber\\
& & -\,\frac{1}{2}\int_Q^T \Lambda^{\phi=0}_{ab,uv}(Q)\bar G^2_{\phi=0}(Q)\bar V^{\phi=0}_{uv,cd}\nonumber\\
& & -\,\frac{1}{2}\int_Q^T \Lambda^{\phi=0}_{ac,uv}(Q)\bar G^2_{\phi=0}(Q)\bar V^{\phi=0}_{uv,bd}\nonumber\\
& & -\,\frac{1}{2}\int_Q^T \Lambda^{\phi=0}_{ad,uv}(Q)\bar G^2_{\phi=0}(Q)\bar V^{\phi=0}_{uv,bc}\nonumber\\
& & -\,\frac{1}{2}\int_Q^T (\Lambda^{\phi=0}_{ab,uv}(Q)-\bar\Lambda^{\phi=0}_{ab,uv})\bar G^2_{\phi=0}(Q)\bar V^{\phi=0}_{uv,cd}\nonumber\\
& & -\,\frac{1}{2}\int_Q^T (\Lambda^{\phi=0}_{ac,uv}(Q)-\bar\Lambda^{\phi=0}_{ac,uv})\bar G^2_{\phi=0}(Q)\bar V^{\phi=0}_{uv,bd}\nonumber\\
& & -\,\frac{1}{2}\int_Q^T (\Lambda^{\phi=0}_{ad,uv}(Q)-\bar\Lambda^{\phi=0}_{ad,uv})\bar G^2_{\phi=0}(Q)\bar V^{\phi=0}_{uv,bc}\,.\nonumber\\
\eeq
Once again $\hat V^{\phi=0}_{abcd}=\hat V_{\phi=0}(\delta_{ab}\delta_{cd}+\delta_{ac}\delta_{bd}+\delta_{ad}\delta_{bc})$, with
\beq
\hat V_{\phi=0} & = & \frac{\lambda_4}{3N}+2\big[V^{(A)}_{\phi=0}-\Lambda^{(A)}_{\phi=0}\big]+4\big[V^{(B)}_{\phi=0}-\Lambda^{(B)}_{\phi=0}\big]\nonumber\\
& & -\,\big[\bar V^{(A)}_{\phi=0}-\bar\Lambda^{(A)}_{\phi=0}\big]-2\big[\bar V^{(B)}_{\phi=0}-\bar\Lambda^{(B)}_{\phi=0}\big],
\eeq
from which we deduce Eq.~(\ref{eq:lambda4_hybrid}).\\

\subsection{Finiteness of ${\cal C}$ and $\tilde {\cal C}$}
Let us show that ${\cal C}$ and $\tilde {\cal C}$ defined in \eqref{eq:calc} and \eqref{Eq:C_tilde} are finite. One possibility is to compute the divergent part of the setting-sun sum integrals and check that ${\cal C}$ and $\tilde {\cal C}$ are free of divergences. As far as ${\cal C}$ is concerned this was done within dimensional regularization in Appendix~B of Ref.~\cite{Marko:2012wc}. We can treat $\tilde {\cal C}$ along similar lines. With the exception of the setting-sun integral ${\cal S}[\bar G_{\rm L};\bar G_{\rm T};\bar G_{\rm T}],$ all the integrals needed to obtain the finite expression of $\tilde {\cal C}$ are given in Ref.~\cite{Marko:2012wc}. Using the method of Ref.~\cite{Blaizot:2004bg}, one can obtain for this setting-sun integral a decomposition in terms of zero, one and two statistical factors analogous to Eq.~(B5) of Ref.~\cite{Marko:2012wc}. From that point on, the calculation of ${\cal S}[\bar G_{\rm L};\bar G_{\rm T};\bar G_{\rm T}]$ and $\tilde {\cal C}[\bar G_{\rm L},\bar G_{\rm T},G_\star]$ parallels that of $S[G]$ and ${\cal C}[G,G_\star]$ performed there and uses the vacuum part of the setting-sun integral with two different masses. For the part with no statistical factors one has to expand the factors of the product ${\cal T}_{\rm F}[\bar G_{\rm}]B_\star[G_\star](0)$ to ${\cal O}(\epsilon)$ because both contain $1/\epsilon$ divergences. Using for ${\cal S}^{(0)}$ the expression given in Sec.~3 of \cite{Davydychev:1992mt} one obtains
\begin{widetext}
\beq
\tilde {\cal C}^{(0)}[\bar G_{\rm L},\bar G_{\rm T},G_\star]&=&\frac{1}{(16\pi^2)^2}\left\{-4m^2_\star+\bar M_{\rm T}^2\left[\left(\ln\frac{\bar M_{\rm T}^2}{m^2_\star}-2\right)^2+\frac{4\pi^2}{9}-\frac{2}{3}\Psi_1\left(\frac{2}{3}\right)-2\Phi(z)\right]\right.
\nonumber\\
&&\left.\qquad\qquad-\bar M_{\rm L}^2\left[\frac{1}{2}\ln^2(4z)-\frac{2\pi^2}{9}+\frac{1}{3}\Psi_1\left(\frac{2}{3}\right)-\frac{1}{2}\Phi(z)\right]
\right\}\,,
\eeq
where $\Psi_1(x)=d^2\Gamma(x)/d x^2$ is the trigamma function, $z=\bar M_{\rm L}^2/(4 \bar M_{\rm T}^2),$ and the function $\Phi(z)$ is defined as
\beq
\Phi(z)=\begin{cases}
\displaystyle 4\sqrt{\frac{z}{1-z}}\,\textrm{Cl}_2(2\arcsin \sqrt{z}),&\text{ if }z<1,\\
\displaystyle \frac{1}{\zeta}\left(-4\,\textrm{Li}_2\left(\frac{1-\zeta}{2}\right)+2\ln^2\left(\frac{1-\zeta}{2}\right)-\ln^2(4z)+\frac{\pi^2}{3}\right),&\text{ if }z>1,
\end{cases}
\eeq
with $\textnormal{Cl}_2(x)=-\int\limits_0^x d\theta\,\ln (2\sin\left(\theta/2\right))$ being the Clausen function and $\zeta(z)=\sqrt{1-1/z}.$ Note that $\displaystyle\lim_{z\to 1} \Phi(z)=8\ln 2.$ The part with one statistical factor reads
\beq
\tilde {\cal C}^{(1)}[\bar G_{\rm L},\bar G_{\rm T},G_\star]&=&\frac{{\cal B}^{(1)}_\star[G_\star](0)}{8\pi^2}
\left[(m_\star^2-\bar M_{\rm T}^2)\left(3-\frac{\pi}{\sqrt{3}}\right)+\bar M_{\rm T}^2\ln\frac{\bar M_{\rm T}^2}{m_\star^2}\right]
+\frac{{\cal T}^{(1)}_\star[G_\star]}{8\pi^2}\left(3-\frac{\pi}{\sqrt{3}}-\frac{\bar M_{\rm T}^2}{m_\star^2}\right)\nonumber\\
&&+F_{\rm L}[\bar M_{\rm L},\bar M_{\rm T}]{\cal T}^{(1)}[\bar G_{\rm L}] 
+2F_{\rm T}[\bar M_{\rm L},\bar M_{\rm T}]{\cal T}^{(1)}[\bar G_{\rm T}]\,, 
\eeq
where 
\beq
F_{\rm L}[\bar M_{\rm L},\bar M_{\rm T}]&=&\frac{1}{16\pi^2}\left[-\ln(4z)-\frac{\pi}{\sqrt{3}} + Q\begin{cases}
\displaystyle \textrm{arctanh}(Q),&\text{ if } z\ge1,\\
\displaystyle \textrm{arctan}\big(Q^{-1}\big),&\text{ if } z<1,
\end{cases}\right],
\\
F_{\rm T}[\bar M_{\rm L},\bar M_{\rm T}]&=&\frac{1}{16\pi^2}\left[\ln\frac{\bar M_{\rm T}^2}{m_\star^2}-2z\ln(4z)-2 + 4zQ\begin{cases}
\displaystyle -\frac{1}{2}\ln\frac{1+Q}{1-Q},&\text{ if } z\ge1,\\
\displaystyle \textrm{arctan}\frac{(2z)^{-1}-1}{Q} + \textrm{arctan}\frac{1}{Q},&\text{ if } z<1,
\end{cases}\right],
\eeq
with $Q=\sqrt{\left|1-1/z\right|}.$ Finally, the part with two statistical factors is 
\beq
\tilde {\cal C}^{(2)}[\bar G_{\rm L},\bar G_{\rm T},G_\star]&=&
2{\cal T}^{(1)}[\bar G_{\rm T}]{\cal B}_\star^{(1)}[G_\star](0)-\frac{1}{3}\left[3{\cal S}^{(2)}[\bar G_{\rm L};\bar G_{\rm T};\bar G_{\rm T}]-{\cal S}^{(2)}[\bar G_{\rm L}]-2{\cal S}^{(2)}_\star[G_\star]-2(\bmt-m_\star^2)\frac{d{\cal S}^{(2)}_\star[G_\star]}{dm_\star^2}\right],\nonumber\\
\eeq
where 
\beq
{\cal S}^{(2)}[\bar G_{\rm L};\bar G_{\rm T};\bar G_{\rm T}]&=&\frac{1}{32\pi^4}\int_0^\infty d p \int_0^\infty d k p k\frac{n_T(\bar \varepsilon_{\rm T}(k))}{\bar \varepsilon_{\rm T}(k)}\left[\frac{n_T(\bar \varepsilon_{\rm T}(p))}{\bar \varepsilon_{\rm T}(p)}\ln\frac{4\varepsilon_{\rm T}^2(k)\varepsilon_{\rm T}^2(p)-(\bar M^2_{\rm L}-2\bar M^2_{\rm T}+2kp)^2}{4\varepsilon_{\rm T}^2(k)\varepsilon_{\rm T}^2(p)-(\bar M^2_{\rm L}-2\bar M^2_{\rm T}-2kp)^2}\right. \nonumber\\
&&\left. +  2\frac{n_T(\bar \varepsilon_{\rm L}(p))}{\bar \varepsilon_{\rm L}(p)}\ln\frac{4\varepsilon^2_{\rm T}(k)\varepsilon^2_{\rm L}(p)-(\bar M_{\rm L}^2-2kp)^2}{4\varepsilon^2_{\rm T}(k)\varepsilon^2_{\rm L}(p)-(\bar M_{\rm L}^2+2kp)^2}\right],
\eeq
\end{widetext}
with $n_T(\varepsilon)=1/(\exp(\varepsilon/T)-1),$ $\varepsilon_{\rm T/L}^2(k)=k^2+\bar M_{\rm T/L}^2,$ and all other integrals are given in Ref.~\cite{Marko:2012wc}.\\

We can also discuss the finiteness of ${\cal C}$ and $\tilde {\cal C}$ using another explicit method, which is less calculational and shows clearly the different roles played by $\lambda_{2{\rm l}}$ and $\delta\lambda_{2{\rm nl}}$. Let us illustrate it in the case of ${\cal C}$. Using the results of Ref.~\cite{Blaizot:2004bg} to write
\beq
{\cal T}[\bar G]={\cal T}_{T=0}[\bar G]+\int_q\frac{n_{\varepsilon_q}}{\varepsilon_q}
\eeq
and
\beq
{\cal S}[\bar G] & = & {\cal S}_{T=0}[\bar G]+3\int_q\frac{n_{\varepsilon_q}}{2\varepsilon_q}\sum_{\sigma=\pm 1}{\cal B}[\bar G](\tilde Q_\sigma)\nonumber\\
& & +\,3\int_q\frac{n_{\varepsilon_q}}{2\varepsilon_q}\int_k\frac{n_{\varepsilon_k}}{2\varepsilon_k}\sum_{\sigma,\tau=\pm 1}\bar G(\tilde Q_\sigma+\tilde K_\tau),
\eeq
with $\int_q\equiv \int d^3q/(2\pi)^3$ and where ${\cal B}[\bar G](\tilde Q_\sigma)$ denotes the analytical continuation of the bubble sum integral to real values of the frequency followed by its evaluation on shell $Q=(i\omega_n,q)\to \tilde Q_\sigma=(q_0=\sigma\varepsilon_q+i0^+,q)$. Plugging this back into the first line of Eq.~(\ref{eq:calc}), we obtain
\beq
{\cal C}[\bar G,G_\star] & = & \frac{2}{\lambda^2_\star}\left(\frac{\lambda_{2{\rm l}}}{\lambda_0}-1\right)(\bar M^2-m^2_\star)+{\rm finite}\nonumber\\
& & +\int_Q^{T=0}\delta \bar G(Q)\left[{\cal B}_\star[G_\star](0)-{\cal B}_\star[G_\star](Q)\right]\nonumber\\
& & +\int_q\frac{n_{\varepsilon_q}}{2\varepsilon_q}\sum_{\sigma=\pm 1}\left[{\cal B}_\star[G_\star](0)-{\cal B}[\bar G](\tilde Q_\sigma)\right]\nonumber\\
& & -\int_q\frac{n^\star_{\varepsilon^\star_q}}{2\varepsilon^\star_q}\sum_{\sigma=\pm 1}\left[{\cal B}_\star[G_\star](0)-{\cal B}_\star[G_\star](\tilde Q_\sigma)\right],\nonumber\\
\label{eq:41}
\eeq
where $\delta\bar G\equiv \bar G-G_\star$, $\varepsilon_q\equiv\sqrt{q^2+\bar M^2}$ and $\varepsilon^\star\equiv\sqrt{q^2+m^2_\star}$. The role of $\delta\lambda_{2{\rm nl}}$ is thus to remove the subdivergences in such a way that the integrands in Eq.~(\ref{eq:41}) are finite. In addition the last two integrals are finite due to the presence of the thermal factors. This is not the case for the zero temperature integral because the factor $\delta \bar G$ does not decrease fast enough in the UV. However the term involving the ratio $\lambda_{2{\rm l}}/\lambda_0$ has precisely the same form as this integral. Separating the $T=0$ part of $\lambda_{2{\rm l}}/\lambda_0$ and combining it with the rest we obtain
\beq
{\cal C}[\bar G,G_\star] & = & \int_Q^{T=0} G_{\rm r}(Q)\left[{\cal B}_\star[G_\star](0)-{\cal B}_\star[G_\star](Q)\right]\nonumber\\
& & +\int_q\frac{n_{\varepsilon_q}}{2\varepsilon_q}\sum_{\sigma=\pm 1}\left[{\cal B}_\star[G_\star](0)-{\cal B}[\bar G](\tilde Q_\sigma)\right]\nonumber\\
& & -\int_q\frac{n^\star_{\varepsilon^\star_q}}{2\varepsilon^\star_q}\sum_{\sigma=\pm 1}\left[{\cal B}_\star[G_\star](0)-{\cal B}_\star[G_\star](\tilde Q_\sigma)\right]\nonumber\\
& & +\,{\rm finite}\,,
\eeq
where $G_{\rm r}\equiv\delta\bar G+(\bar M^2-m^2_\star)G_\star^2=(\bar M^2-m^2_\star)^2G_\star^2\bar G$ decreases fast enough to make the corresponding integral convergent. A similar proof can be given for $\tilde {\cal C}.$ 

%%%
\subsection{Explicitly finite field equation}
We give here the explicitly finite form of the field equation. Differentiating Eq.~\eqref{eq:hyb_effpot_N} with respect to $\phi$ and taking into account the implicit dependence on $\phi$ of the gap masses one obtains
\beq
&&\bar\phi\bigg\{\bar M^2_{\rm L}-\frac{\lambda_\star}{3N}\bar\phi^2+\frac{\lambda_\star^2}{18 N^2} \bigg[{\cal C}_N[\bar G_{\rm L},\bar G_{\rm T},G_\star]
\nonumber\\\nonumber\\
&&+\bar\phi^2\Big((N+8) {\cal D}[\bar G_{\rm L},G_\star]+(N-1)\tilde {\cal D}_{\rm L}[\bar G_{\rm L},\bar G_{\rm T}]\Big) \frac{d \bar M_{\rm L}^2}{d\bar\phi^2}
\nonumber\\
&&
+\bar\phi^2 (N-1)\tilde {\cal D}_{\rm T}[\bar G_{\rm L},\bar G_{\rm T},G_\star]\frac{d \bar M_{\rm T}^2}{d\bar\phi^2} \bigg]\bigg\} = 0,\ \ 
\label{Eq:hybrid_EoS}
\eeq
where, in order to give a compact expression, we have already used in the first two terms, coming from the Hartree-Fock part, the solution of the two linear equations for the derivatives of the gap masses which are obtained from Eqs.~\eqref{eq:bmlfinite} and \eqref{eq:bmtfinite} as
\beq
\frac{d \bar M_{\rm L}^2}{d\bar\phi^2}=\frac{d-b}{a d - c b},\qquad \frac{d \bar M_{\rm T}^2}{d\bar\phi^2}=\frac{a-c}{a d - c b},
\eeq 
with  $c={\cal B}[\bar G_{\rm L}](0)-{\cal B}_\star[G_\star](0),$ $a=c+2N/\lambda_\star,$ $b=(N-1)\big[{\cal B}[\bar G_{\rm T}](0)-{\cal B}_\star[G_\star](0)\big]/3,$ and $d=3b(N+1)/(N-1)+6N/\lambda_\star.$ Both $b$ and $c$ are finite because the divergence of the perturbative bubble integral does not depend on the mass. We note that for $N=1$ one has $b=0$ and $d\bar M_{\rm L}^2/d\bar\phi^2=1/a,$ therefore with the notation $\bar M^2\equiv \bar M_{\rm L}^2$ the field equation simplifies to
\beq
0&=&\bar\phi\bigg[\bar M^2-\frac{\lambda_\star}{3}\bar\phi^2+\frac{\lambda_\star^2}{2} {\cal C}[\bar G,G_\star]\nonumber\\
&&+\frac{\lambda_\star^2}{4}\bar\phi^2\frac{{\cal D}[\bar G,G_\star]}{\lambda_\star^{-1}+\big[{\cal B}[\bar G](0)-{\cal B}_\star[G_\star](0)\big]/2}\bigg].
\label{Eq:hybrid_EoS_N=1}
\eeq 

In Eq.~\eqref{Eq:hybrid_EoS}, ${\cal C}_N$ is defined in Eq.~\eqref{Eq:C_N} and ${\cal D}[\bar G_{\rm L},G_\star],$ whose explicit expression is given in Eq.~(B12) of Ref.~\cite{Marko:2012wc}, is obtained from ${\cal C}[\bar G_{\rm L},G_\star]$, defined in \eqref{eq:calc}, upon differentiation with respect to $\bar M^2_{\rm L}.$ Finally, $\tilde {\cal D}_{\rm L}[\bar G_{\rm L},\bar G_{\rm T}]$ and $\tilde {\cal D}_{\rm T}[\bar G_{\rm L},\bar G_{\rm T},G_\star]$ are obtained from $\tilde {\cal C}[\bar G_{\rm L},\bar G_{\rm T},G_\star]$ given in Eq.~\eqref{Eq:C_tilde} upon differentiation with respect to $\bar M_{\rm L}^2$ and $\bar M_{\rm T}^2,$ respectively.\\

In order to determine the $\bar\phi_{\rm c,0}$ curve in the $(m_\star^2/T_\star^2,\lambda_\star)$ plane, discussed below Eq.~\eqref{Eq:Phi_c_T}, we have to take the limit $T\to0$ and $\bar M_{\rm T}\to 0$ in Eq.~\eqref{Eq:hybrid_EoS}. This is straightforward, except for the last term where $\tilde {\cal D}_{\rm T}$, which at $T=0$ comes only from $\tilde {\cal C}^{(0)}$, diverges and $d\bar M_{\rm T}^2/d\bar\phi^2$ vanishes in this limit. Working out the limit one obtains
\beq
&&\lim_{\bar M^2_{\rm T}\to 0} \left[\tilde {\cal D}_{\rm T}[\bar G_{\rm L},\bar G_{\rm T},G_\star] \frac{d \bar M_{\rm T}^2}{d \phi^2}\right]\bigg|_{T=0}\nonumber\\
&&=\frac{\frac{1}{8\pi^2}\left(\ln\frac{\bar M_{\rm L,0}^2}{m_\star^2}-1\right)+2{\cal B}^{(1)}_\star[G_\star](0)}{\frac{(N+2)\lambda_\star}{3N}\left({\cal B}^{(1)}_\star[G_\star](0)+\frac{1}{16\pi^2}\ln\frac{\bar M_{\rm L,0}^2}{m_\star^2}\right)-N-1}.\qquad
\eeq

As a last remark, note that there is no equivalent of the $\bar\phi_{\rm c,0}$ line in the $N=1$ case, since this requires $\bar\phi_0>0$ and $\bar M_0=0,$ and by examining Eq.~\eqref{Eq:hybrid_EoS_N=1} at $T=0$ one sees that the two conditions cannot be satisfied simultaneously. This is because, for $\bar M_0\to 0$, ${\cal C}$ is finite, but ${\cal D}$ diverges as $\ln^2(\bar M_0/m_\star),$ while the denominator diverges only as $\ln(\bar M_0/m_\star),$ meaning that in this limit $\bar\phi_0=0$ is the only solution of the field equation.

%%%
\subsection{Solution of the gap equation at $\phi=0$ and $T=0$\label{ss:Lambert}}
\begin{figure}[!t]
\begin{center}
\includegraphics[width=0.48\textwidth,angle=0]{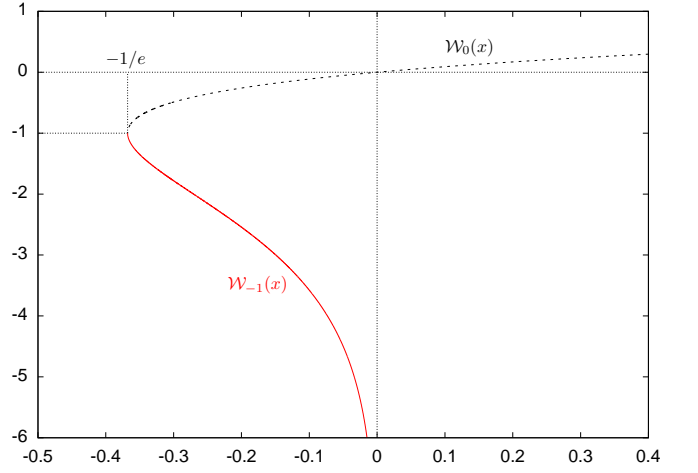}
\caption{The two real branches of the Lambert ${\cal W}$ function. The upper branch ${\cal W}_0(x)$ (dashed) is defined for $x\in[-1/e,\infty)$ and the lower one ${\cal W}_{-1}(x)$ (solid) for $x\in[-1/e,0).$ \label{Fig:Lambert-W}}
\end{center}
\end{figure}

We show here that the solution(s) of Eq.~\eqref{Eq:bmz} can be given at $T=0$ in terms of the two real branches of the Lambert ${\cal W}$ function defined to be the multivalued inverse of the function $w\mapsto we^w=z$, for $w$ complex. This function verifies ${\cal W}(z)\exp({\cal W}(z))=z$ for any complex $z.$ The real branches of the Lambert ${\cal W}$ function are depicted in Figure~\ref{Fig:Lambert-W}. The upper branch is usually called ${\cal W}_0(x)$ and the lower one ${\cal W}_{-1}(x).$  

At $T=0$ one can rewrite Eq.~\eqref{Eq:bmz} as
\beq\label{Eq:bmzz}
\bar M_{\phi=0,T=0}^2\ln\bigg(e^{b_\star/a_\star}\frac{\bar M_{\phi=0,T=0}^2}{m^2_\star}\bigg) = - \frac{C_\star}{a_\star},
\eeq
where $a_\star=(N+2)\lambda_\star/(96\pi^2 N),$ $b_\star=-1+(N+2)\lambda_\star\big[{\cal B}_\star^{(1)}[G_\star](0)-1/(16\pi^2)\big]/(6N)$ and $C_\star$ is defined
in Eq.~\eqref{Eq:Cstar}. With a few algebraic manipulations (exponentiation and multiplication by $-C_\star/(a_\star \bar M_{\phi=0,T=0}^2)$) and using the definition of the Lambert function, one expresses the solution of Eq.~\eqref{Eq:bmzz} as
\beq\label{Eq:M_Labert}
\bar M_{\phi=0,T=0}^2=-\frac{C_\star/a_\star}{\displaystyle {\cal W}\bigg(-\frac{C_\star}{m_\star^2a_\star}e^{b_\star/a_\star}\bigg)}.
%\vspace*{0.6cm}\\\nonumber
\eeq
For $C_\star>0$ (points below the $\bar T_{\rm c}=0$ line of Figure~\ref{Fig:chiralParams}) the argument of ${\cal W}$ is negative and one sees by looking at Figure~\ref{Fig:Lambert-W} that for $\bar M_{\phi=0,T=0}$ one has no solution if $-\frac{C_\star}{m_\star^2a_\star}e^{b_\star/a_\star}<-1/e$ and two solutions if $-\frac{C_\star}{m_\star^2a_\star}e^{b_\star/a_\star}<-1/e,$ one smaller and one bigger than $\bar M_{\rm e}= 2 \Lambda_{\rm p}^{\rm est}/e,$ where $\Lambda_{\rm p}^{\rm est}$ is the accurate estimation of the Landau pole given in \eqref{Eq:Lp_approx} (the two solutions merge when  $C_\star e^{b_\star/a_\star}=m_\star^2a_\star/e$). The lower scale solution is given by the lower branch ${\cal W}_{-1}$ of the Lambert function.\footnote{We know that at $\bar T_{\rm c}=0$ one has $\bar M_{\phi=0,\bar T_{\rm c}=0}=0,$ and this can be obtained only with ${\cal W}_{-1},$ which diverges negatively when its argument vanishes. The use of the other branch would give a finite value, because ${\cal W}_0(x)=x + {\cal O}(x^2),$ for small $x.$} The larger scale solution is given by the upper branch ${\cal W}_0.$ For $C_\star\le 0$ one has one solution, bigger than $\bar M_{\rm e}$ and given by ${\cal W}_0.$ 

In conclusion, to obtain the $T_{\rm c}=0$ curve, defined as $\hat M_{\phi=0,T_{\rm c}=0}=0$, we have to take the solution \eqref{Eq:M_Labert} given by the lower branch of the Lambert function and use it in Eq.~\eqref{Eq:lc_line}, which can be solved only numerically. For small negative arguments ${\cal W}_{-1}$ can be evaluated using the asymptotic series given in Ref.~\cite{Lambert_Wm1}. We mention finally that the upper branch plays a role because we have considered the renormalized gap equation in its continuum limit. If we would consider it in the presence of a finite 3D cutoff, the solutions corresponding to the upper branch would be absent, see Ref.~\cite{Reinosa:2011ut}.\\

%%%%%
\section{Tensor decomposition}\label{app:tensor}
Let us first consider a real symmetric tensor $U^\phi_{ab}=U^\phi_{ba}$ such that for any rotation $R\in SO(N)$
\beq\label{eq:C1}
U^{R\phi}_{ab}=R_{ac}R_{bd}U^\phi_{cd}\,.
\eeq
The case $\phi=0$ is easily treated. For any $\lambda\in\mathds{R}$ we have
\beq
(U^{\phi=0}_{ab}-\lambda\delta_{ab})R_{bc}=R_{ab}(U^{\phi=0}_{bc}-\lambda\delta_{bc})\,.
\eeq
Since the fundamental representation of $SO(N)$ is irreducible, it follows from Schur's lemma that $U-\lambda\mathds{1}$ is either $0$ or invertible. If we chose $\lambda$ to be an eigenvalue of $U^{\phi=0}$ (there exists at least one eigenvalue since $U^{\phi=0}$ is real and symmetric), $U^{\phi=0}-\lambda\mathds{1}$ cannot be invertible and thus $U^{\phi=0}=\lambda\mathds{1}$. For reasons that will appear below, the case $\phi\neq 0$ requires that we distinguish $N=2$ from $N>2$. Let us consider the case $N>2$ first. Since $U^\phi$ is real and symmetric, it is diagonalizable, that is it admits $N$ linearly independent eigenvectors. Let us consider an eigenvector $u^\phi$ which is not collinear to $\phi$ (such an eigenvector exists for $N>2$). If $R$ is a rotation that leaves $\phi$ invariant, we have from (\ref{eq:C1}):
\beq
U^\phi_{ab}R_{bc}u^\phi_c=R_{ab}U^\phi_{bc}u^\phi_c=\lambda^\phi R_{ab}u^\phi_b\,,
\eeq
which shows that we have indeed at least $N-1$ linearly independent eigenvectors corresponding to the eigenvalue $\lambda^\phi$. If the remaining eigenvector corresponds also to $\lambda^\phi$, we have $U^\phi=\lambda^\phi\mathds{1}$ with $\lambda^{R\phi}=\lambda^\phi$ that is with $\lambda^\phi$ a function of $\phi^2$ only. If the remaining eigenvector corresponds to another eigenvalue $\mu^\phi\neq\lambda^\phi$, it has to be collinear to $\phi$. If it were not, we could construct $N-1$ linearly independent eigenvectors, different from the previous ones and it would follow that $N\geq 2N-2$ that is $N\leq 2$, which contradicts our assumption $N>2$. From this and the fact that the subspaces are orthogonal to each other, it follows that
\beq\label{eq:C4}
U^\phi_{ab}=\mu^\phi P^{\rm L}_{ab}+\lambda^\phi P^{\rm T}_{ab},
\eeq
with $\lambda^\phi$ and $\mu^\phi$ functions of $\phi^2$ only. The case $N=2$ is particular because the solution to Eq.~(\ref{eq:C1}) is much more general than (\ref{eq:C4}). In fact, because $SO(2)$ is Abelian, any tensor of the form 
\beq
U^\phi_{ab}=\tilde R_{ac}\tilde R_{bd}\left(\mu^\phi P^{\rm L}_{cd}+\lambda^\phi P^{\rm T}_{cd}\right),
\eeq
with $\tilde R\in SO(2)$, obeys Eq.~(\ref{eq:C1}). The converse can also be proven to be true. By imposing that Eq.~(\ref{eq:C1}) holds not only for any $R\in SO(2)$ but also for any $R\in O(2)$, one recovers the form (\ref{eq:C4}).

%%%%%

\end{document}